\documentclass[useAMS,usenatbib]{mn2e}
\usepackage[usenames,dvipsnames]{color}
\usepackage{lastpage}
\usepackage{float}  
\usepackage{titlesec}
\def\mnras{MNRAS}
\def\apj{ApJ}
\def\aap{A\&A}
\def\araa{ARA\&A}
\newcommand{\ssr}{Space~Sci.~Rev.}	
\newcommand{\apjl}{ApJLett}		
\newcommand{\pasp}{PASP}		
\newcommand{\aj}{AJ}			
\newcommand{\pasa}{PASA}			
\newcommand{\nat}{Nature}		
\newcommand{\bain}{Bull. Astron. Inst.}		
\def\na{New A}
\newcommand{\apjs}{ApJS}		
\newcommand{\Msun}{\hbox{$\hbox{M}_\odot\;$}}
\newcommand{\Rsun}{\hbox{$\hbox{R}_\odot\;$}}
\newcommand{\myemail}{s.repetto@astro.ru.nl}
\newcommand{\beq}{\begin{equation}}
\newcommand{\eeq}{\end{equation}}
\newcommand{\mr}{\mathrm}
\usepackage{amssymb,amsmath}
\usepackage{epsfig}
\usepackage{rotating}
\usepackage{times}
\usepackage{natbib,twoopt}

\usepackage[breaklinks=true, colorlinks=true]{hyperref} 
\hypersetup{colorlinks=True, linkcolor=BlueGreen}
\bibpunct{(}{)}{;}{a}{}{,}             
\makeatletter
  \newcommandtwoopt{\citeads}[3][][]{\href{http://adsabs.harvard.edu/abs/#3}%
    {\def\hyper@linkstart##1##2{}%
     \let\hyper@linkend\@empty\citealp[#1][#2]{#3}}}
  \newcommandtwoopt{\citepads}[3][][]{\href{http://adsabs.harvard.edu/abs/#3}%
    {\def\hyper@linkstart##1##2{}%
     \let\hyper@linkend\@empty\citep[#1][#2]{#3}}}
  \newcommandtwoopt{\citetads}[3][][]{\href{http://adsabs.harvard.edu/abs/#3}%
    {\def\hyper@linkstart##1##2{}%
     \let\hyper@linkend\@empty\citet[#1][#2]{#3}}}
  \newcommandtwoopt{\citeyearads}[3][][]%
    {\href{http://adsabs.harvard.edu/abs/#3}
    {\def\hyper@linkstart##1##2{}%
     \let\hyper@linkend\@empty\citeyear[#1][#2]{#3}}}
\makeatother

\title[Short-period black-hole low-mass X-ray binaries]
  {Constraining the formation of black-holes in short-period
    Black-Hole Low-Mass X-ray Binaries}
\author[S.~Repetto and G.~Nelemans]
  {Serena~Repetto$^{1}$\thanks{E-mail:
\myemail},~~Gijs~Nelemans$^{1,2}$\\
    $^1$Department of Astrophysics/IMAPP, Radboud University Nijmegen, P.O. Box 9010, 6500 GL Nijmegen, The Netherlands\\
    $^2$ Institute for Astronomy, KU Leuven, Celestijnenlaan 200D, 3001 Leuven, Belgium
  }

\date{\today}

\pagerange{\pageref{firstpage}--\pageref{}} \pubyear{0000}
\def\LaTeX{L\kern-.36em\raise.3ex\hbox{a}\kern-.15em
    T\kern-.1667em\lower.7ex\hbox{E}\kern-.125emX}

\begin{document}

\label{firstpage}

\maketitle

\begin{abstract}
The formation of stellar mass black holes is still very uncertain.
Two main uncertainties are the amount of mass ejected in the supernova
event (if any) and the magnitude of the natal kick the black hole
receives at birth (if any).  {{Repetto et al. (2012)}}, studying the
position of Galactic X-ray binaries containing black holes, found
evidence for black holes receiving high natal kicks at birth.  In this
Paper we extend that study, taking into account the previous binary evolution
of the sources as well.  The seven short-period black-hole X-ray
binaries that we use, are compact binaries consisting of a low-mass
star orbiting a black hole in a period less than $1$ day.  We trace
their binary evolution backwards in time, from the current observed
state of mass-transfer, to the moment the black hole was formed, and
we add the extra information on the kinematics of the binaries. We
find that several systems could be explained by no natal kick, just
mass ejection, while for two systems (and possibly more) a high kick is required. So
unless the latter have an alternative formation, such as within a
globular cluster, we conclude that at least some black holes get high
kicks. This challenges the standard picture that black hole kicks
would be scaled down from neutron star kicks. Furthermore, we find
that five systems could have formed with
a non-zero natal kick but zero mass ejected (i.e. no supernova) at
formation,
as predicted by neutrino-driven natal kicks.

\end{abstract}


\section{Introduction}
Since the discovery of the first stellar-mass black hole (BH) in the
Galactic X-ray source Cygnus X-1 \citep{1965Sci...147..394B},
other BHs have been found.  Stellar-mass\footnote{Throughout
  this Paper, we will refer to stellar-mass BHs simply as
  BHs.} BHs have a mass between $\sim5$ and $30$ times the
mass of the Sun, while the peak of the Galactic distribution is
centered at $8~\Msun$ \citep{2010ApJ...725.1918O}. 
{{Their Galactic
population currently amounts to $17$ dynamically confirmed BHs
\citep{2014SSRv..183..223C},
and $31$ BH candidates (Tetarenko, B.E. et al. 2015, in prep).}}

So far, the best way to detect BHs is when they are actively
accreting from a stellar companion, whose matter falls in the
gravitational field of the BH while forming an accretion disc.
{{We wish to mention the recent discovery of a BH 
with a Be-type star as companion (\citealt{2014Natur.505..378C}).
This system, in which the accretion flow onto the BH is radiatively inefficient,
opens-up the possibility of a new detection-window for BHs.}}
In black-hole low-mass X-ray binaries (BH-LMXBs), a BH
accretes matter from a star similar in mass to our Sun. At some point
in the evolution of the progenitor of a BH-LMXB, the companion
overfills its Roche lobe, either because of its own nuclear expansion,
or due to shrinking of the orbit caused by angular momentum losses.
The material hence escapes the gravitational pull from the star
and forms an accretion disc around the BH,
which is detectable in the X-ray band \citep{1973A&A....24..337S}.

Despite the strong evidence for the existence of BHs, the way
they actually form is still a matter of great debate.  It is generally
accepted that BHs are formed from the gravitational collapse
of a massive star, a star with a mass on the main sequence (MS)
greater than $20-25~\Msun$, and/or a core mass greater than $8~\Msun$,
e.g. \citet{1999ApJ...522..413F}, \citet{2001ApJ...550..410M}, \citet{2006csxs.book..623T}.
In one scenario, massive BHs are thought to form by direct collapse,
whereas the lightest ones via fallback onto the nascent neutron star
(NS). Two main uncertainties are the size of the velocity
the BH receives at birth (\emph{natal kick}, NK), if any,
and the amount of mass ejected at the moment of BH formation,
if any. These are the questions we address in this Paper.

Measuring BH natal kicks is of great importance for a number
of reasons. First of all, the magnitude of the NK is dependent 
upon the physical mechanism driving the kick. For instance,  if BHs were discovered
to receive high kicks, this would have important implications
for the supernova (SN) mechanism. Theoretical calculations by
\citet{2013MNRAS.434.1355J} show that the momentum of the BH can grow with the fallback
mass and lead to high kicks. Secondly, the velocity that BHs receive
at birth affects the number of BHs which can be retained in a globular
cluster (GC, \citealt{2012Natur.490...71S}, \citealt{2013MNRAS.430L..30S}, 
\citealt{2015ApJ...800....9M}), {{as well as in a young stellar cluster (\citealt{2014ApJ...781...81G})}}. Furthermore, the black hole NK
distribution is an important ingredient in binary population synthesis
(BPS) studies, for examples the ones which compute the gravitational
wave (GW) merger rate of binaries harboring BHs (see for example \citealt{1997MNRAS.288..245L}
and \citealt{2015ApJ...806..263D}). Additionally, the likelihood of discovering isolated (and nearby) BHs accreting
from the interstellar medium (ISM) through radio signatures detectable
by future surveys,
is a function of the BH velocity relative to the ISM,
hence of the NK (\citealt{2013MNRAS.430.1538F}).

In our first study of BH natal kicks (\citealt{2012MNRAS.425.2799R}),
we found evidence for BHs receiving {\emph{high}} kicks,
{{where {\emph{high}} stands for 
kicks similar to those received by NSs (that is,
hundreds of km/s)}}. In that study, high NKs were needed in
order to match the current out-of-the-plane distribution of BH X-ray
binaries.

As currently observed binary properties
(e.g., orbital period, masses, and position in the Galactic potential)
are determined by the conditions present at the moment of BH formation,
and therefore significantly affected by the magnitude of both
the ejected mass and NK imparted,
BH-LMXB systems are the optimal tool for shedding light on BH formation 
mechanisms.

In this Paper, we focus on a subset of BH-LMXBs, those ones with a
short orbital period (less than $1$ day).  In these systems the
current mass transfer phase is driven by angular momentum loss in the
form of magnetic braking (MB).  This study is complementary to the study of
\citet{2012MNRAS.425.2799R} (Paper I hereafter).  
Contrary to Paper I,
which used a BPS approach to study the NK magnitude
required to explain the current out-of-plane distribution of 
Galactic BH-LMXBs,
in this Paper we use a detailed binary evolution approach,
allowing us to take into account the current orbital parameters of
each system. We trace the binary evolution of
the seven short-period BH-LMXBs backwards in time until the moment the
BH was formed. This allows us to set constraints on the mass
ejected and the NK in terms of lower limits.

Previous works on BH natal kicks combined the study of the
binary evolution of the systems with the information on the space
velocity. A small NK has been found for
GRO J1655-40 (\citealt{2005ApJ...625..324W}) 
{{and
for Cygnus X-1 (\citealt{2012ApJ...747..111W})}}, whereas evidence for a
high NK was found for XTE J1118+480
(\citealt{2009ApJ...697.1057F}).  {{Evidence for a BH formed in a
SN explosion comes from the chemical enrichment in the spectra
of the companion to the BH in XTE J1118+480
(\citealt{2006ApJ...644L..49G}), in GRO J1655-40
(\citealt{1999Natur.401..142I}), in 1A 0620-00 (\citealt{2004ApJ...609..988G}), in V4641 Sgr (\citealt{2001ApJ...555..489O}),
and in V404 Cyg (\citealt{2011ApJ...738...95G}). 
{{Whereas Cygnus X-1 was claimed
to have formed without a SN (\citealt{2003Sci...300.1119M}).}}}}

The Paper is structured as follows: in Section~\ref{sec:overview} we
give a brief overview of the evolution of BH-LMXBs and 
of what we know about BH formation, in
Section~\ref{sec:Galpop} we show the properties of the observed
systems. In Section~\ref{sec:method} we describe our method to
calculate the NK and mass ejected for the seven binaries, and in
Section~\ref{sec:results} the results. We end with a discussion and
conclusion of our findings.

\section{A short overview of the evolution of BH-LMXBs}
\label
{sec:overview}
The standard formation scenario for a BH-LMXB is an extension of the formation scenario
which explains low-mass X-ray binaries with a NS as accretor (\citealt{1983adsx.conf..308V}).
The primordial binary from which a neutron-star low-mass X-ray binary originates, is a binary with a very extreme mass ratio between the two components.
In order to form a BH low-mass X-ray binary, one just needs to start with a more massive progenitor
which will core-collapse into a BH, instead of a NS (\citealt{1996IAUS..165...93R}, \citealt{1997A&A...321..207P}, \citealt{1998A&A...338...69E}).
The progenitor of a BH low-mass X-ray binary probably consists of a Sun-like
star orbiting around a massive star (with a mass of $20-25~M_\odot$).
Typical orbital separations which allow a sun-like star to overfill
its Roche lobe and transfer mass to a BH are of the order of $\sim
10~R_\odot$. These separations are much smaller than the typical
orbital separation that the progenitor of the BH-LMXB must have had in
order to accommodate the progenitor of the BH during its nuclear
expansion. Therefore, it is commonly accepted that the binary must
have undergone a phase of common envelope (CE), which shrank the
binary down by a factor of $\sim100$ (\citealt{1976IAUS...73...75P}).  During the CE, the massive star
is believed to lose its H-envelope; what is left, the Helium core,
collapses into a BH.  If the binary survives the event in which the BH
forms, its further evolution can follow two alternative paths.
The evolution can be driven by the nuclear evolution of the stellar component, 
and the binary will undergo mass transfer when the star is large enough for 
some of its material to fall in the gravitational potential of the BH.
At the current state, the companion star is thus an evolved star off the MS.
Alternatively, the evolution can be driven by angular momentum losses from the binary
and the two components get closer and closer to each other thanks
to the coupled effect of {{tides}} and magnetic braking, until mass transfer sets in.
The latter ones are the binaries we focus on in this study,
where the companion star is a MS-star. For previous studies of the evolution
of BH-LMXBs see e.g. 
\citet{1998A&A...338...69E},
\citet{1999ApJ...521..723K},
\citet{2003MNRAS.341..385P},
\citet{2006A&A...454..559Y}.

\subsection{Black hole formation and natal kicks at birth}
In the most common scenario of BH formation, higher mass BHs are
thought to form via direct collapse of the progenitor star, lighter
mass BHs via fallback onto the proto-NS
(\citealt{2001ApJ...554..548F}, \citealt{2001ApJ...550..410M},
\citealt{2003ApJ...591..288H}, 
\citealt{2008ApJ...679..639Z}).  BH formation via fallback occurs in a
successful but weak explosion, where some fraction of the ejecta does
not have enough kinetic energy to escape the potential well of the
proto NS. However, we wish
to highlight that the occurrence of fallback is not agreed upon by the
entire SN community.  Some studies showed that the fallback
requires a fine-tuning between the explosion energy and the envelope
binding energy (see for example \citealt{2010MNRAS.405.2113D}). Normally,
either the SN is successful and leads to the formation
of a NS, or it is unsuccessful and all the material collapses into a
BH. The BH mass would therefore be equal to the mass of the
collapsing Helium core (see \citealt{2015MNRAS.446.1213K} and \citealt{2015ApJ...799..190C}).

A further source of uncertainty is whether BHs receive NKs at
birth or not, and what the size of the NK would be.  There are two
types of BH natal kicks,
kicks (i) imparted intrinsically to the BH,
and (ii) received as an effect of the NK imparted to the NS that forms a BH through fallback.
The intrinsic kicks could be caused either by asymmetric gravitational
wave emission during the core-collapse
(\citealt{1995MNRAS.273L..12B}), or by asymmetric flux of those
neutrinos which get to escape before all the collapsing material falls
inside the event horizon (\citealt{1993A&A...271..187G}). In either
cases, there is no need for mass-ejection at BH formation.  

There are two main processes which are thought to cause NS
natal kicks and are therefore relevant for BH formation via fallback:
either asymmetries in the SN ejecta ({\emph{ejecta-driven}}, also
called {\emph{hydrodynamical}} NKs), or asymmetries in the
neutrino flux ({\emph{neutrino-driven}} NKs).  Both of 
these two NK mechanisms have their own drawbacks (\citealt{2006ApJS..163..335F}). Only $1\%$ of the collapse
energy is in the ejecta, thus the hydrodynamics of the explosion has
to develop a large degree of anisotropy to impart the NS a
large NK. Instead, most of the energy and momentum liberated
in the explosion are carried away by neutrino, hence only a small
degree of asymmetry ($\sim 1\%$) in the neutrino flux is necessary to
impart the NS a NK as high as hundreds
km/s. However, this mechanism requires a strong magnetic field.

If the BH is formed via fallback onto the proto NS, the magnitude of
the NK it receives depends on the competition between two timescales:
the timescale for (some of) the material in the ejecta to fall back
onto the nascent NS, $\tau_\mr{fb}$, and the timescale for
the process which imparts the NS a NK,
$\tau_\mr{NK}$. If $\tau_\mr{fb}>\tau_\mr{NK}$ we expect the BH to
receive a reduced NK.  If $\tau_\mr{fb}<\tau_\mr{NK}$, we expect the
BH to receive a full NK. While hydrodynamic simulations
by \citet{2001ApJ...550..410M}
show that fallback onto the NS happens within $100$ s after core collapse,
the timescale of NKs induced by neutrino
emission 
and by hydrodynamics in the ejecta
is estimated to be $\tau_\mr{NK}\sim 10$ s and $\tau_\mr{NK}\sim 0.1$ s respectively \citep{2001ApJ...549.1111L}.
This indicates that, in the fallback scenario, the BH would receive a reduced NK.

Just by conserving linear momentum, we expect reduced NKs to
be of the order of $V_\mr{NK, BH} \sim \left ( M_\mr{NS}/M_\mr{BH}
\right )V_\mr{NK, NS}$, leading the BH to receive a NK of the order of
tens km/s.  If this scenario is correct, we should find a correlation
between the mass of the BH and the NK: the larger the BH mass, the
larger the fallback mass, the lower the NK.
We wish to highlight, however, that
\citet{2013MNRAS.434.1355J} suggests BHs could be
accelerated to the same velocity as NSs, 
even in a fallback-scenario, due to the
anisotropic gravitational pull from the  asymmetrically expelled ejecta. Thus, in this scenario,
BHs would receive high kicks even when $\tau_\mr{fb}>\tau_\mr{NK}$,
and the momentum of the BH would grow with the fallback mass.

For NSs there is both evidence for high and low
kicks. The proper motion of isolated pulsars implies transverse
velocities up to $1000$ km/s, with a mean birth speed for young
pulsars of $\sim 400$ km/s (\citealt{2005MNRAS.360..974H}). 
On the other hand, few Galactic NS high-mass X-ray binaries
seem to be more consistent with a low NK.
These binaries have long periods
and small eccentricities. 
Such a small eccentricity,
being a prior of the eccentricity immediately after the SN explosion,
 requires a small NK, of the order of few tens km/s (\citealt{pfahl2002b}).
High NS natal kicks are thought to be produced in a standard
core-collapse SN, whereas low NS natal kicks are thought to be
produced in a less-energetic type of SN, an electron-capture SN from a
low-mass star (\citealt{2004ApJ...612.1044P}).
Such a low-mass star
has little chance of forming a BH. When the collapsing core is low in mass, the explosion is thought to be so rapid that there is no time
for large asymmetries to develop. Such a theoretical scenario is consistent with the results obtained tracing backwards the 
orbital properties of a few members of the Galactic population of
double NS binaries (\citealt{2010ApJ...721.1689W}). Few of them are consistent with a low-mass progenitor
and a low NK at birth. The retention of
NSs in GCs is another argument in favor of some NSs receiving low kicks (\citealt{pfahl2002a}).

\section{The Galactic population of short-period black-hole low-mass X-ray binaries}
\label{sec:Galpop}

\subsection{Orbital properties}
\label{sec:OrbProp}
In our Galaxy there are $13$ BH-LMXBs {{with BH masses dynamically measured}}
(\citealt{2004ApJ...616..376O},
\citealt{2006csxs.book..157M}). {{Dynamical measurement of the BH mass
is done tracing the Keplerian motion of the companion star.
In this way, the BH mass can be determined once
the orbital period, the radial velocity semi-amplitude of the companion star,
the mass-ratio, and the inclination of the binary with respect to the line of sight, are known.
The BH mass is a strong function of the inclination,
hence it is the inclination the major contributor to any uncertainties in the BH mass.
The inclination is typically determined fitting the ellipsoidal modulations in the optical light curve 
of the companion.
The companion
can be detected when the binary
{{is in quiescence}},
i.e. when the accretion disc is under-luminous. 
Failure in accurately determining the contribution of the disc to the light curve, 
results {{in inaccurate}} BH mass determination.
For an up-to-date overview of dynamical mass determinations in BHs,
and the associated main sources of systematic errors,
see \citet{2014SSRv..183..223C}.}}

\begin{figure}
\centering
\includegraphics[width=0.9\columnwidth]{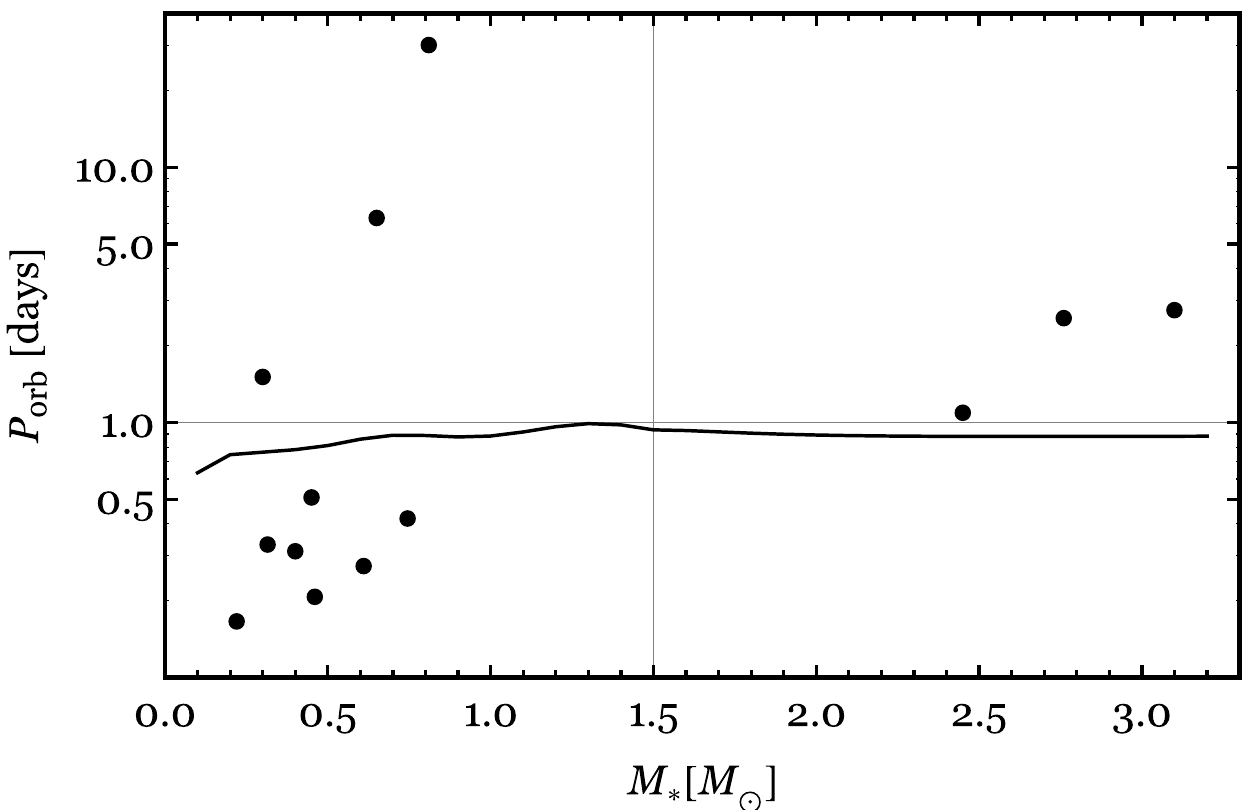}
\caption{Black-hole low-mass X-ray binaries as a function of orbital period $P_\mr{orb}$ and companion mass $M_\star$. The solid line shows the bifurcation period. The sources in the bottom left of the plot are those we study in this work.}
\label{fig:PorbVsM2}
\end{figure}

In Fig.~\ref{fig:PorbVsM2} we show
the orbital
period $P_\mr{orb}$ of BH-LMXBs as a function of the mass of the companion $M_\star$. They fall in three categories: $10$ short-period
($P_\mr{orb}\lesssim 1$ day) systems, of which $7$ with $M_\star\lesssim 1.5~M_\odot$,
and $3$ systems with longer periods. Out of this sample, we only
consider the short-period binaries with low-mass companions. We don't consider the binary with $1.5$ day period as a short-period
binary, since it contains a giant as a companion
(\citealt{2011ApJ...730...75O}), so it is thought to evolve to longer
periods. 

The
choice $P_\mr{orb}\lesssim 1$ day is motivated by the fact that
binaries with longer periods are driven by the evolution of the donor,
while only those with shorter periods are driven by magnetic braking
and evolve towards smaller and smaller periods until the companion
overfills its Roche lobe, while still being on the MS (see
\citealt{2006A&A...454..559Y}). This defines a so-called bifurcation
period $P_\mr{bif}$ (\citealt{1985SvAL...11...52T}, \citealt{1989A&A...208...52P}), that is shown as the solid line in
Fig. \ref{fig:PorbVsM2}. We calculate $P_\mr{bif}$ as the the orbital period such that the
magnetic braking time scale ($\tau_\mr{MB}$, calculated below in Section~\ref{subsec:coupling}) is
equal to the main-sequence life-time. The limit $M_\star\lesssim
1.5~M_\odot$ is motivated by the fact that more massive donors are not
thought to sustain a significant surface magnetic field, hence
magnetic braking is not supposed to operate.
{{We wish to mention that \citet{2006MNRAS.366.1415J}
proposed that BH-LMXBs could descend from a population of intermediate-mass X-ray binaries
with Ap/Bp-star companions. However, the spectral types predicted by the stellar evolutionary tracks
are too hot for explaining the observed temperatures of companion stars in BH-LMXBs.}}

We indicate the dynamical
properties of the seven short-period BH X-ray binaries in Table \ref{tab:tabOBSpro}.
{{We wish to note that GRS 1009-45 lacks strong constraints on the mass of the BH.
This is due to the large uncertainty on the inclination of the system,
and the BH-mass value given in \citet{1999PASP..111..969F} is a lower-limit
(see \citealt{2014SSRv..183..223C}).
Another BH-LMXB whose observational properties are debated is GRO J0422+32.
\citet{2007MNRAS.374..657R} claimed that 
negligible
contamination arising from the accretion disc was assumed in the estimate from 
 \citet{2003ApJ...599.1254G}. They estimate the contamination to the companion-star light curve due to the accretion disc,
 and they get a lower limit to the BH mass of $\sim 10.4~M_\odot$.
 {{Although both the reported masses are probably biased,
 because of very large flickering amplitude in the light curves
(\citealt{2014SSRv..183..223C}),
we find the measurement by  \citet{2003ApJ...599.1254G} as the most reliable one,
 because it is based on a database where the ellipsoidal modulation is best detected.
 Neglecting the disc contribution to the light curve would, in any case,
 underestimate the inclination, and consequently overestimate the BH mass,
 in strong contrast with the results of \citet{2007MNRAS.374..657R}.}}
}}

\begin{table}
\caption{{Dynamical properties of short-period black-hole low-mass X-ray binaries.}}
\label{tab:tabOBSpro}
\begin{tabular}{l c c c c}
\hline
Source& $P_\mr{orb}$ & $M_\star$& $M_\mr{BH}$ & Ref.\\
& \small{[day]} & \small{[$M_\odot$]} & \small{[$M_\odot$]} & \\
\hline\hline
XTE J1118+480 & 0.17 & 0.15-0.29 & 8.16-8.58 & [1] \\
GRO J0422+32 & 0.21 & 0.15-0.77 & 3.02-4.92 & [2]\\
GRS 1009-45 & 0.28 & 0.61$^a$ & 4.4$^{a}$ & [3] \\
1A 0620-00 & 0.32 & 0.39-0.41 & 6.44-6.84 & [4] \\
GS 2000+251 & 0.34 & 0.16-0.47 & 5.5-8.8 & [5] \\
Nova Mus 91 & 0.43 & 0.68-0.81 & 6.35-7.55 & [6] \\
H 1705-250 & 0.52 & 0.3-0.6 & 6.4-6.9 & [7] \\
\hline \\
\end{tabular}
\newline
{{(a) This system lacks strong constraints on the component masses (see Text).}}\\
References: [1] \citet{2012ApJ...744L..25G}, [2] \citet{2003ApJ...599.1254G}, 
[3] \citet{1999PASP..111..969F}, [4] \citet{2014MNRAS.438L..21G}, 
[5] \citet{2004AJ....127..481I}, [6] \citet{2001AJ....122..971G},
[7] \citet{1997PASP..109..461F}.
\end{table}

\subsection{Mass-radius relation}
\begin{figure}
\centering
\includegraphics[width=0.9\columnwidth]{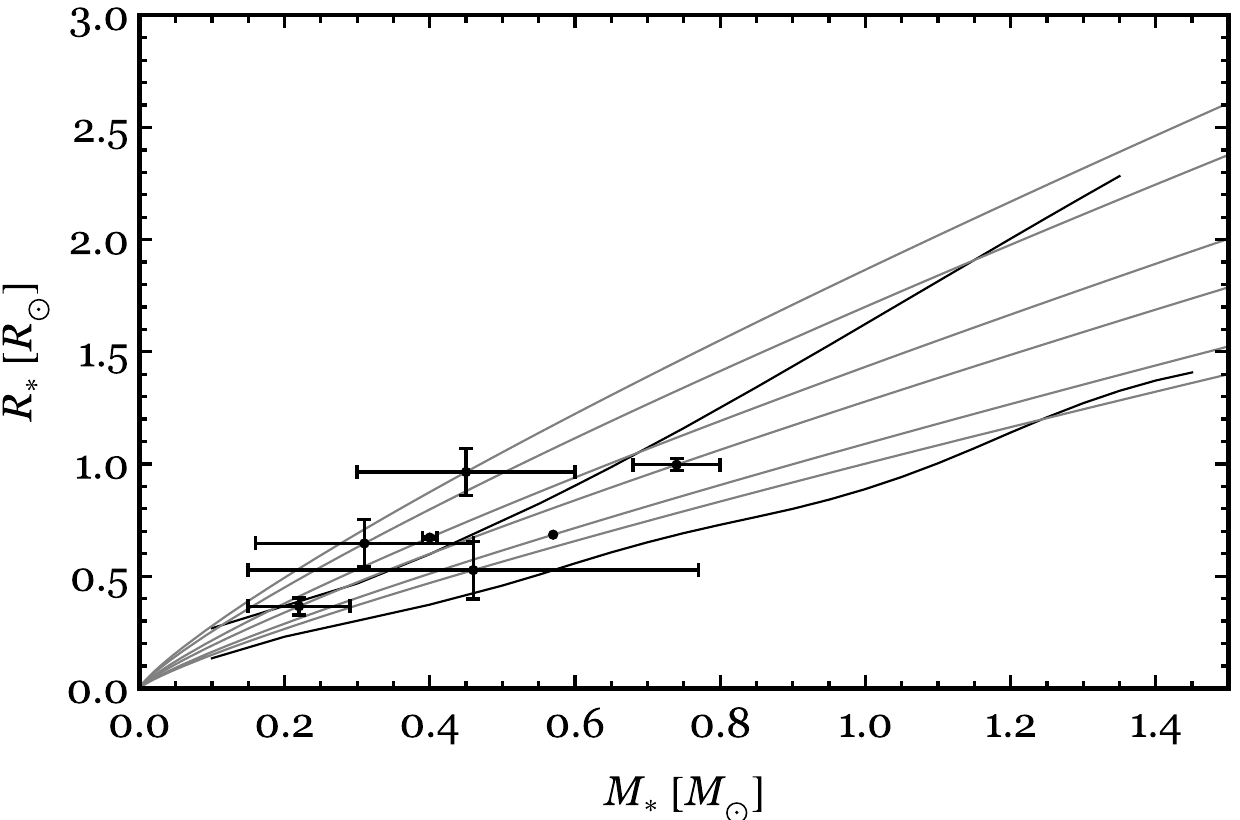}
\caption{Radius and mass of the seven short-period black-hole low-mass X-ray binaries. The two black lines correspond to the radius of single stars at the beginning and at the end of the main sequence. The grey lines correspond to our fit of the mass-radius relation of the seven short-period BH low-mass X-ray binaries.}
\label{fig:R2vsM2}
\end{figure}

For our sample, the companion star is currently overfilling its Roche
lobe. From the observed binary properties and inferred masses we can
hence calculate the radius of the companion assuming $R_\star=R_\mr{L}$, where $R_\mr{L}$
is the Roche lobe radius of the companion star
(\citealt{1983ApJ...268..368E}). The results are plotted in
Fig. \ref{fig:R2vsM2} and can be compared to the radii of single main
sequence stars shown as the solid lines.
We compute such radii for single stars, both at the beginning and at the end of the MS,
with the SSE code by \citet{2000MNRAS.315..543H} embedded in the Astrophysics Multipurpose Software Environment AMUSE (\citealt{2009NewA...14..369P}).

As is well known, stars in interacting binaries are typically somewhat larger
than single stars. {{This can be caused 
by the fact that the star does not have time to relax back to its thermal equilibrium (\citealt{2011ApJS..194...28K}).
In  case one component is a NS or BH
the effect of X-ray irradiation,
which heats up (hence bloats) the star, can play a role
(\citealt{1991Natur.350..136P}, \citealt{1991ApJ...383..739H}).
However,
this effect is not thought to be significant in LMXBs in quiescent state.}}

In order to calculate the evolution of the
binaries, it is needed to know their mass-radius relation,
since we need to know how the star responds to the loss of mass during mass transfer.
For
interacting binaries with white dwarfs accretors (cataclysmic
variables, CVs), it has been shown that a power-law fit to the
mass-radius relation $R_\star\propto M_\star^\alpha $ is an adequate
description (\citealt{2011ApJS..194...28K}).

We therefore model the mass radius relation of the companion stars in
BH-LMXBs as $\frac{R_\star}{R_\odot}=f\left (\frac{M_\star}{M_\odot}\right
)^\alpha$.  We thus assume the BH-binaries to all align on the same
slope in the logM-logR diagram.  We fit the short-period binaries with
a mass-radius index of $\sim 0.82$. We use this common power-law index as a
fixed parameter to fit the mass-radius relation for each system, that
is, for finding $f$ for each binary.
In Fig. \ref{fig:R2vsM2} we show such a common power-law index
with the grey lines, to be compared with the mass-radius relation of single MS stars (solid black lines).
For comparison, the mass-radius relation for CVs has been fitted by \citet{2011ApJS..194...28K} with $\alpha_\mr{CV}=0.69$.

\subsection{Kinematical properties and Galactic positions}
\label{subsec:kin}

The final (and crucial) piece of information we will use to
reconstruct the birth of the BH is {{the}} position of the
binary in the Galaxy. Assuming that
the binary is born in the Galactic plane and receives a kick in the SN
explosion right perpendicular to the plane, 
it will then move on a straight line
reaching a maximum height $z_\mr{max}$.
Assuming that the binary's 
current height $z$ above
the plane is $z_\mr{max}$,
this height
{{gives}} a minimum peculiar velocity at birth $v_{\perp\mr{,min}}$
given the Galactic potential at the distance $R$ from the Galactic
center (Paper I).  
The velocity $v_{\perp\mr{,min}}$ is the minimum velocity at birth
required for bringing the system
from the plane to its current position above (or below) the plane. 
This velocity is calculated
simply by
conserving the energy along the trajectory in the Galactic potential:
$$v_{\perp\mr{,min}}=\sqrt{2[\Phi\left (R_0, z \right ) - \Phi\left ( R_0, 0
    \right )]}, $$ where $\Phi \left (R, z \right )$ is a model for
the Galactic potential (\citealt{1990ApJ...348..485P}) and 
$R_0$ is the observed projected distance of the binary
from the Galactic centre.

\begin{figure}
\centering
\includegraphics[width=0.9\columnwidth]{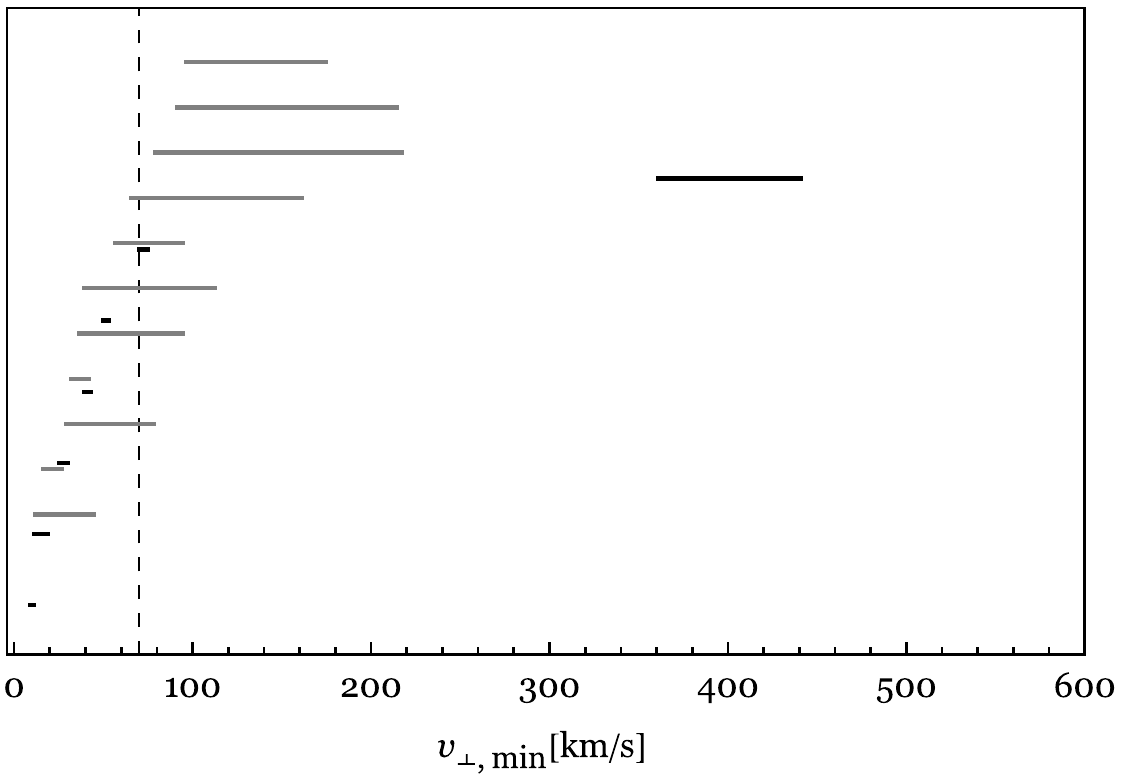}
\caption{Range of possible values for the minimum peculiar velocity at birth of the short-period black hole X-ray binaries from Table \ref{tab:tabkin}.
The solid lines indicate the systems belonging to our study.
The grey lines indicate the BH candidates from Tetarenko, B.E. et al. 2015, in prep. {{The dashed vertical line
indicates the lower limit on the peculiar velocity at birth for XTE J1118+480.}}}
\label{fig:plottrattini}
\end{figure}

In Table
\ref{tab:tabkin} we show the resulting values for $v_{\perp\mr{,min}}$ for our
seven systems, as well as for the tentative BH candidates from Tetarenko, B.E. et al. 2015 (in prep).  These BHs do not have a dynamical BH-mass estimate, but
are likely BHs due to their spectral and timing properties (Tetarenko, B.E.,
priv. comm.). Of the sources contained in Tetarenko et al. catalogue,
we select only the short-period ones. We show
the range of possible values of $v_{\perp,\mr{min}}$ for every source in Fig. \ref{fig:plottrattini}.
It is interesting to note that there
are $5$ other sources whose minimum $v_\perp$
is very similar to the one of XTE J1118+480,
which is one of the sources in our sample consistent with a high NK
(see Sec. \ref{sec:discussion}).
\citet{2013A&A...552A..32K} noted the similarities between XTE J1118+480,
MAXI J1659-152 and
Swift J1753.5-0127,
pointing out their similar orbital period (few hrs)
and the large scale height from the Galactic plane,
suggesting that all three were kicked out of the Galactic plane,
or possibly born in a GC (see also \citealt{2013MNRAS.434.2696S}). We will suggest a similar interpretation
for the remaining high peculiar velocity sources.

An estimate of the actual (rather than minimal) peculiar velocity
received at birth can be obtained if the current 3D space velocity of
the binary is measured. Integrating the trajectory of the binary backwards in time
from the current initial conditions of the position and 3-D velocity,
it is possible to infer the 3-D velocity at birth. 
The only system in our sample with
measured both radial velocity and proper motion, and hence 3-D space velocity, is XTE J1118+480
(\citealt{2014PASA...31...16M}).
Its Galactic trajectory has been integrated backwards 
by \citet{2009ApJ...697.1057F},
giving a peculiar velocity at birth of
$110-240$ km/s,
which is consistent with our lower limit $v_{\perp\mr{,min}}\sim 70$ km/s.
\begin{table}
\caption{{Kinematical properties of short-period black-hole low-mass X-ray binaries. The first group contains the binaries of this study.
The second group contains the putative black hole candidates.}}
\label{tab:tabkin}
\begin{tabular}[h]{l c c c c }
\hline
Source& distance & $z$ & $v_{\perp\mr{,min}}$ & Ref.\\
& \small{[kpc]} & \small{[kpc]} &\small{[km/s]}  & \\
\hline\hline 
XTE J1118+480 & 1.62-1.82 & 1.43-1.61  & 70-75  & [1] \\
GRO J0422+32 & 2.19-2.79 & -[0.45-0.57] & 25-30 & [2]\\
GRS 1009-45 & 3.55-4.09 & 0.58-0.66 & 39-43 & [3] \\
1A 0620-00 & 0.94-1.18 & -[0.11-0.13] & 9-11  & [4] \\
GS 2000+251 & 2-3.4 &-[0.10-0.18] & 11-19  & [5] \\
Nova Mus 91 & 5.63-6.15 & -[0.69-0.76] &  50-53 & [6] \\
H 1705-250 & 6.5-10.7 & 1.02-1.68 & 361-441  & [7] \\
\hline 
MAXI J1305-704	& 2-8 & -[0.26-1.06] & 29-78 & [8]\\ 
Swift J1357.2-0933 & 0.5Ð6.3	& [0.38-4.83] &  91-214  & [9]\\
XTE J1650-500  & 1.9-3.3 &  -[0.11-0.20]  &  16-26 & [10] \\
MAXI J1659-152	 & 4.9-12.3   &  [1.39-3.50] & 79-217 & [11]\\
GRS 1716-249 	 & 2-2.8  &   [0.24	-0.34]  & 32-42 & [12]\\
Swift J174510.8-262411	 &  2-8   & [0.05-0.19] &   12-44 & [8]\\ 
Swift J1753.5-0127   &  2-8 	& [0.42-1.69]  &  66-161 & [8]\\ 
H 1755-338	 & 4-9  &  -[0.34-0.77] &   96-174 & [13]\\
MAXI J1836-194	 & 4-10 & -[0.37-0.93] & 39-112 & [14]\\
XTE J1859+226  & 11-14 &   [1.65-2.09] & 56-94  & [15]\\
4U 1957+115 & 2-8 & -[0.32-1.30]  & 36-94 & [8]\\ 

\hline 
\end{tabular}
\newline
References: [1] \citet{2006ApJ...642..438G}, [2] \citet{2003ApJ...599.1254G}, [3] \citet{2001PhDT..........G},
[4] \citet{2010ApJ...710.1127C}, [5] \citet{1996ApJ...473..963B}, 
[6] \citet{2005ApJ...623.1026H},
[7] \citet{1996ApJ...459..226R},
[8] No acceptable estimates for the distance is available, so
a distance of $5\pm 3$ is assumed as done by Tetarenko, B.E. et al. 2015, in prep., [9] \citet{2013MNRAS.434.2696S},
[10] \citet{2006MNRAS.366..235H}, [11] \citet{2013A&A...552A..32K}, [12] \citet{1994A&A...290..803D},
[13] \citet{1985MNRAS.216.1033M}, [14] \citet{2014MNRAS.439.1390R}, [15] \citet{2011MNRAS.413L..15C}.
\end{table}      

\section{A semi-analytical method to follow the evolution of short-period BH-LMXBs}
\label{sec:method}

In order to link the observed properties of BH-LMXBs
with BH formation, we will have to take the evolution of the
binary since the BH formation into account. We use a semi-analytical
approach to simulate the evolution of each binary in our sample.  We
follow the evolution of the three stages the binary goes through (BH
formation, detached phase, mass transfer), but in reverse order.

Tracing backwards in time the mass transfer phase, we obtain the
orbital properties of the binary at the onset of the Roche Lobe Overflow (RLO), $a_\mr{RLO},
M_{\star, \mr{RLO}}, M_{\mr{BH,RLO}}$,
{{where $a_\mr{RLO}$ is the orbital separation.}} These properties will be a
function of the mass transferred to the BH and of the mass-radius
relation of the companion, both parameters that we vary. We then use
these to calculate the orbital evolution due to the coupled effect of
tides and magnetic braking during the detached phases following the BH
formation. This then defines a set of potential progenitor binaries
and BH formation properties (NK and mass ejection) that
are consistent with the currently observed properties of BH-LMXBs.

\subsection{Mass Transfer}
The evolution of long-period BH binaries is driven by the nuclear
expansion of the donor.  Assuming the orbital angular momentum is conserved,
one can trace the mass-transfer backwards in terms of the mass
transferred to the BH (see \citealt{2009MNRAS.394.1440M}).  We use a
similar approach for dealing with the mass-transfer in short-period
binaries, modeling it in an analytical way.

During Roche Lobe Overflow, the BH companion loses mass and its
radius shrinks accordingly. Assuming that some type of orbital angular momentum loss always ensures the star is
filling its Roche lobe, the shrinkage of the star induces a
consequent shrinkage of the orbit. How much the orbit shrinks depends
on the mass-radius relation $R_\star \sim M_\star^\alpha$ of the star.  A
similar approach was used for following the evolution of CVs by
\citet{2011ApJS..194...28K}.

The rate at which the logarithm of the orbital separation $a$
changes due to the combined effect of mass transfer, mass and angular
momentum loss from the binary, is given by the following balance
equation:
\begin{equation}
\frac{\dot{a}}{a}=2\frac{\dot{J}_\mr{orb}}{J_\mr{orb}}-2\frac{\dot{M}_\mr{BH}}{M_\mr{BH}}-2\frac{\dot{M_\star}}{M_\star}+\frac{\dot{M}}{M},
\label{eq:balance}
\end{equation}
where $M$ is the total mass of the binary and $J_\mr{orb}$ the orbital angular momentum.
We reparametrize the balance equation \ref{eq:balance} in terms of
$M_\star$, eliminating time as a variable.  Solving the corresponding
equation, we obtain the orbital separation $a$ and mass of the BH at
any point in the past, as a function of $M_\star$ and $\alpha$, when
the mass transfer is conservative (i.e. $\dot{M}_\mr{BH}=-\dot{M}_\star$).  When the mass transfer is non
conservative, an additional parameter is $\beta$
($\dot{M}_\mr{BH}=-\beta \dot{M_\star}$),
and
we assume that the matter leaves the binary with the specific angular momentum of the BH.
The resulting analytic formulae are
\begin{equation}
\frac{a}{a_\mr{cur}}=\left ({\frac{M_\star}{M_{\star,\mr{cur}}}}\right )^{-\frac{1}{3}+\alpha},
\label{eq:MBCMT}
\end{equation}
for conservative mass transfer, and 
\begin{equation}
\frac{a}{a_\mr{cur}}=\left ({\frac{M_\mr{BH}}{M_{\mr{BH},\mr{cur}}}}\right )^{2-\frac{2}{\beta}}\left ({\frac{M_\star}{M_{\star,\mr{cur}}}}\right )^{-\frac{1}{3}+\alpha}\left (\frac{M}{M_\mr{cur}}\right )^{-\frac{8}{3}},
\label{eq:MBNCMT}
\end{equation}
for non-conservative mass transfer. The $\mr{cur}$-subscript indicates values
at the current time.

The mass transfer is traced backwards in this way until the onset of
mass transfer. In order to define this, we have to assume how much
mass is transferred since the onset, $\Delta M$, and this
defines $M_{\star, \mr{RLO}} = M_{\star, \mr{cur}} + \Delta M$. In our
standard model we assume $\Delta M = 1~M_\odot$.
We will vary our assumptions on the mass transfer,
namely the amount of transferred mass and the 
 mass-radius index, when performing our simulations in Section~\ref{sec:results}.
 A smaller transferred mass and/or a smaller mass-radius index,
 will make the orbital separation at the onset of RLO smaller.

\subsection{Detached evolution and BH formation}

\subsubsection{Coupling between tides and magnetic braking}
\label{subsec:coupling}

To calculate the evolution during the detached phase preceding the
mass transfer, we need to model the evolution of a system in which both tides and magnetic braking operate. 
Magnetic braking is the loss of angular momentum in a magnetic stellar wind
(see \citealt{1958ApJ...128..677P}, \citealt{1967ApJ...148..217W}, \citealt{1981A&A...100L...7V}).
We refer to
\citet{2014MNRAS.444..542R} for details on the numerical method used
to take into account the coupling between tides and magnetic braking.
{{In short, 
after BH formation,
the tidal torque circularizes and synchronizes the binary.
From here, every bit of angular momentum lost from the star in its magnetic wind
is also lost from the orbit,
effectively shrinking the binary until the onset of RLO.}}

Our test-bed MB prescription was introduced by \citet{1981A&A...100L...7V},
and it gives the rate at which the star spins down $\dot{\omega}_\star$.
This in turn gives the MB timescale, $\tau_\mr{MB}=J_\mr{orb}/\dot{J}_\star=J_\mr{orb}/I_\star\dot{\omega}_\star$,
where $I_\star=k^2 M_\star R_\star^2$ is the moment of inertia of the star.
Assuming the star is synchronized with the orbit ($\omega_\star=\omega_\mr{orb}$), one gets:
\begin{equation}
\tau_\mr{MB}= \frac{M_\mr{BH}}{M^2} \frac{1}{G \gamma_\mr{MB} k^2}\frac{a^5}{R_\star^4},
\end{equation}
where $k$ is the gyration radius of the star
and $\gamma_\text{MB}$ is measured as $\approx 5\times 10^{-29} \mr{s/cm}^2 $.

For each of the seven binaries, we compute $a_\mr{max}$ such that RLO happens
within the MS lifetime of the BH companion. The maximal orbital separation $a_\mr{max}$ is a
function of the eccentricity of the binary right after the BH
formation, and it depends on the MB calibration. The value
of $a_\mr{max}$ varies only very little when taking a different
calibration factor for tides, as shown in
\citet{2014MNRAS.444..542R}. 

We numerically fit $a_\mr{max}$ as a function of the
eccentricity $e$ and of the orbital separation at the onset of RLO,
$a_\mr{RLO}$:
\begin{equation}
a_\mr{max}(e, a_\mr{RLO})=\alpha+\beta(1-\exp({\gamma e})) +\frac{\delta e}{e-\epsilon}
\label{eq:fitamax}
\end{equation}
where $\alpha, \beta, \gamma, \delta, \epsilon$ are the parameters of the fit,
and they depend on $a_\mr{RLO}$ (see Appendix \ref{sec:appC}).  \\

We
test two types of magnetic braking law: one from
\citet{1981A&A...100L...7V} (VZ, hereafter), and the other one from
\citet{2003ApJ...599..516I} (IT, hereafter).  The tidal model is
based on \citet{1981A&A....99..126H} and we use calibration factors
as in \citet{2002MNRAS.329..897H}.

\begin{figure}
\centering \includegraphics[width=0.9\columnwidth]{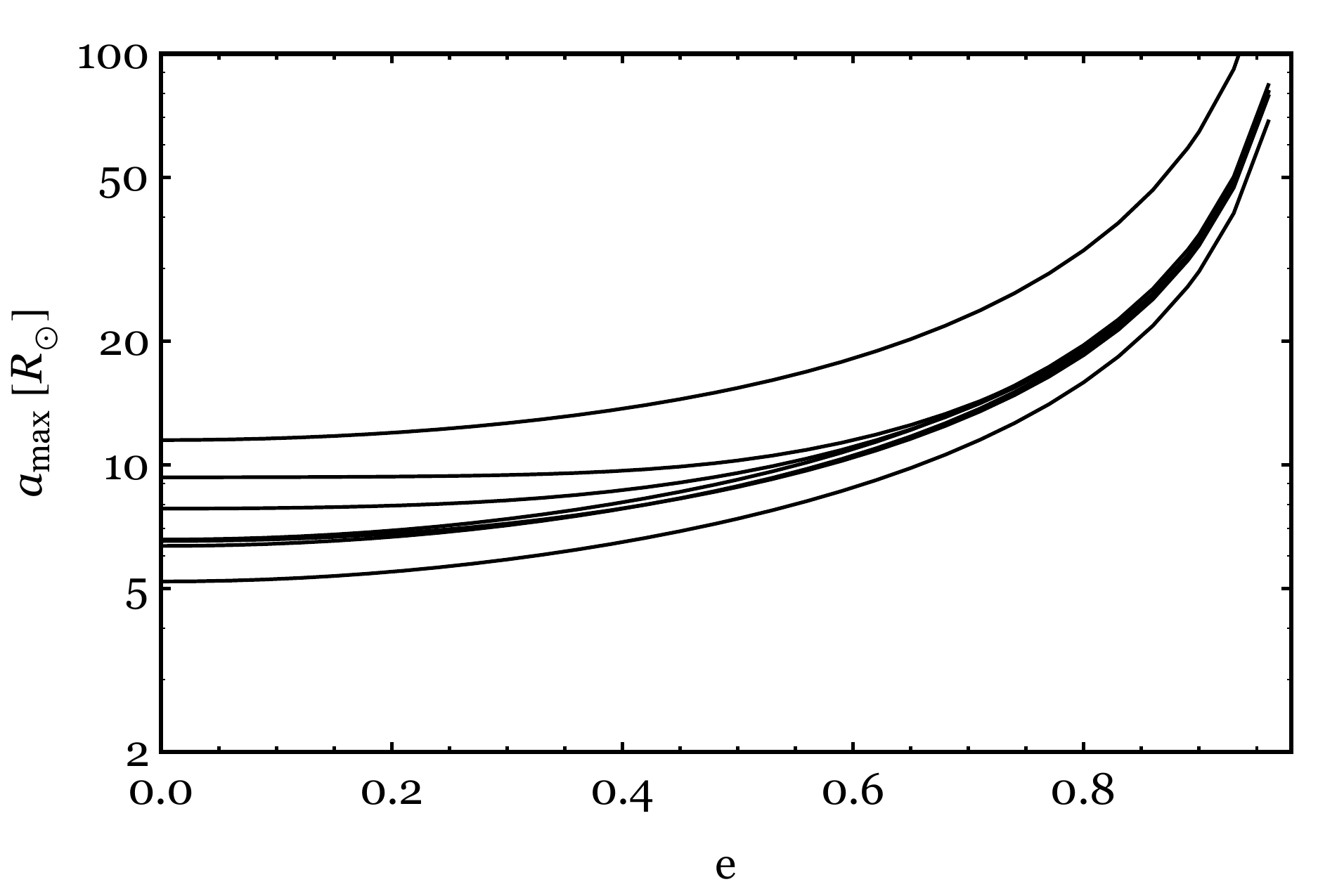}
\caption{Maximal orbital separation right after black hole formation so that the star in short-period black-hole low-mass X-ray binaries fills its Roche Lobe within the main-sequence lifetime.
From bottom to top, the curves are associated to the following sources: GRO J0422+32, 1A 0620-00 and Nova Mus 91 (overlapping curves),
GRS 1009-45, GS 2000+251, H 1705-250, XTE J1118+480.}
\label{fig:amaxvsecc}
\end{figure}

We show $a_\mr{max}$ as a function of the eccentricity in Fig. \ref{fig:amaxvsecc} for the seven sources we are studying,
when using a VZ-type of magnetic braking. The orbital separation of the RLO configuration $a_\mr{RLO}$
is computed using formula \ref{eq:MBCMT},
tracing the mass transfer phase backwards in time from the current properties,
until the companion-star mass increases to $1~M_\odot$.

We wish to note that the orbital separation $a_\mr{max}$ is not affected by the rate at which the companion star is initially spinning.
This is due to the fact that the angular momentum stored in the star
is much less than the angular momentum stored in the orbit.
The ratio between the value of $a_\mr{max}$ when taking a star
initially spinning at $\omega_\star \approx 0.9~\omega_\mr{break}$
and the value when the stellar spin is $\omega_\star \approx 10^{-6}~\mr{s}^{-1}$,
is between $1-1.5$ for every value of the eccentricity.

Another way of computing $a_\mr{max}$
is simply by requiring that $\tau_\mr{MB}\leq \tau_\mr{MS}$,
as done for instance by \citet{1999ApJ...521..723K}.
The orbital separation used for computing $\tau_\mr{MB}$
is the circularized orbital separation after BH formation.
As an example, 
in the case of the binary XTE J1118+480,
$a_\mr{max}$ computed using timescale considerations
differs only little from $a_\mr{max}$ in our model, by $15\%$ at most.

\subsection{Supernova dynamics}
\label{subsec:SN}
The orbital configuration right after BH
formation \footnote{Throughout the whole paper, we denote with
  \emph{-pre} the binary configuration right before the BH formation,
  with \emph{-post} the configuration right after.}  can be determined
univocally from the properties of the pre-BH-formation configuration
through conservation of orbital energy and orbital angular momentum.
An abundance of earlier studies deal with the effect of a SN explosion
and possibly of a NK on the orbital configuration
(see for example \citealt{1961BAN....15..265B}, \citealt{1961BAN....15..291B},
\citealt{1983ApJ...267..322H}, \citealt{1996ApJ...471..352K}).  In
this way, one can compute the orbital elements $a_\mr{post},
e_\mr{post}$ knowing $a_\mr{pre}$, $e_\mr{pre}$, the magnitude and
direction of the NK, and the amount of mass ejected from the
collapsing star (mass which is assumed to leave the system
instantaneously without further interacting with the binary). The
effect of the NK and the mass ejected at BH formation combine
together to give the total peculiar (also called {\emph{systemic}}) velocity of the binary:
\begin{equation}
V_{\rm pec} = \sqrt{\left(\frac{M_{\rm BH}}{M^\prime}\right)^2 V_{\rm NK}^2 + V_{\rm MLK}^2 -2\frac{M_{\rm BH}}{M^\prime}V_{\rm NK,x}V_{\rm MLK}},
\label{eq:PEC}
\end{equation}
where $M^\prime$ is the total mass after the binary has lost a mass $M_\mr{ej}$ in the SN,
$V_{\rm NK}$ is the magnitude of the NK,
$V_{\rm NK,x}$ its component along the orbital speed of the BH progenitor,
and $V_{\rm MLK}$ is the {\emph{mass-loss kick}},
the recoil the binary gets after a mass $M_\mr{ej}$ is lost,
{{also known as {\emph{Blaauw kick}} (\citealt{1961BAN....15..265B})}}:
\begin{equation}
V_{\rm MLK} = \frac{M_\mr{ej}}{M^\prime} \frac{M_\star}{M} \sqrt{{\frac{GM}{a}}},
\label{eq:massloss}
\end{equation}
{{where $M$ is the total mass of the binary right before the SN.}}

\begin{figure}
\includegraphics[width=0.9\columnwidth]{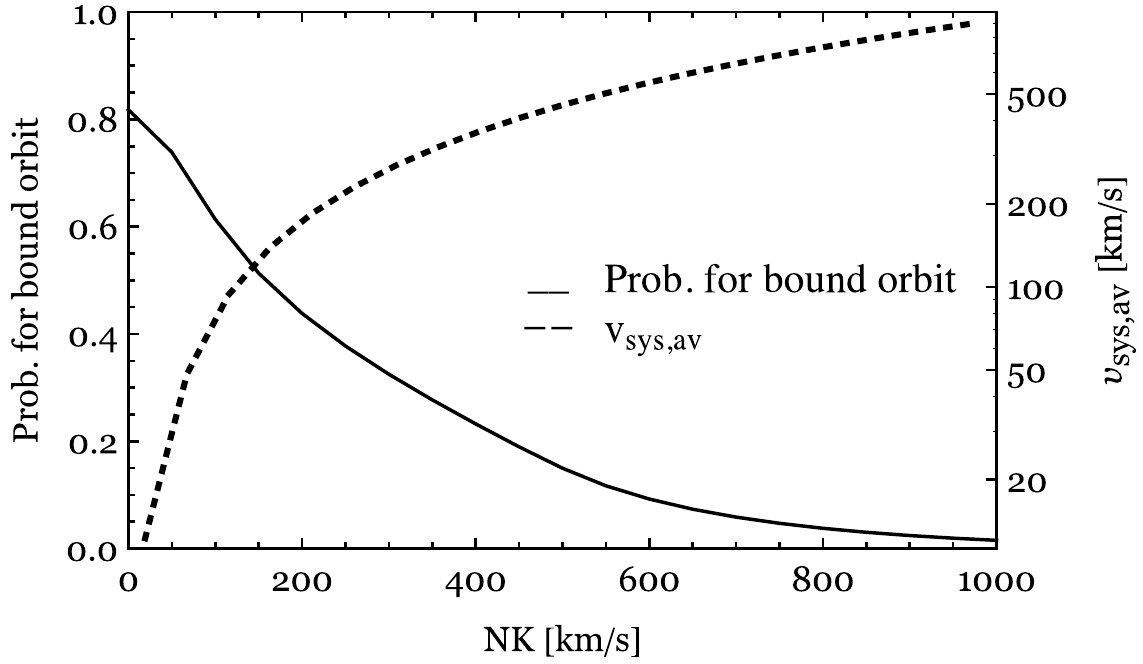}
\caption{Probability for a binary containing a black hole of mass $8~M_\odot$ and a companion with mass $1~M_\odot$
of having survived the supernova event 
 (solid line) as a function of the natal kick. The dashed line shows
 the averaged systemic velocity acquired by the 
  binary in the supernova event.}
\label{fig:boundprob}
\end{figure}

In Fig. \ref{fig:boundprob}, we show the probability for the binary
to stay bound in the SN event (solid line). This probability is calculated for a
sample of binaries formed by a BH of mass $8~M_\odot$ and a companion
of mass $1~M_\odot$, with randomized initial orbital separations and
ejected mass, and random orientation of the NK. The probability is
calculated as a function of the NK. We also show (dashed line) the
average systemic velocity acquired by the binary in case it stays
bound.

\subsection{Monte Carlo calculation of BH formation and detached evolution}

Once we have the properties at RLO, we treat the BH formation phase
with a Monte Carlo approach.  We build a sample of $80\times 10^6$
initial binaries formed by a companion of mass $M_{\star, \mr{RLO}}$
and the progenitor of the BH with mass
$M_\mr{prog}=M_{\mr{BH,RLO}}+M_\mr{ej}$, where $M_\mr{ej}$ is drawn
from a uniform distribution between $0$ and $10~M_\odot$. The orbital
separation of the initial binary is drawn from a uniform distribution
between a minimum value and $50~R_\odot$, and the orbit is assumed to
be synchronous and circular. The minimum value for the orbital
separation is the value at which the Helium star and/or the companion
overfill their Roche lobe. The Helium star ejects a mass equal to
$M_\mr{ej}$, and what is left collapses into a BH, which receives a NK
at formation drawn from a uniform distribution between $0$ and $1000$
km/s. The inclination of the NK with respect to the orbital plane is
uniformly peaked over a sphere centered on the progenitor of the BH.

The motivation which lays behind our choice of the range for the
initial orbital separation is statistics.  Orbital separations larger
than $50~\Rsun$ will only very rarely lead to RLO within the MS
lifetime. Specifically, RLO on the MS will happen only if the binary
is highly eccentric in the post-SN configuration. This can be seen
from the decaying PDFs at large values for the pre-SN orbital separation in
Fig. \ref{fig:aprepre}. We tested this assumption checking how many systems with initial orbital separation
$a_\mr{pre}>50~R_\odot$ evolve into successful ones. For all sources except XTE J1118+480,
no simulated binary with $a_\mr{pre}>50~R_\odot$ evolves into a successful one. For XTE J1118+480
the fraction of systems which evolve into successful ones is $9\%$.

Concerning the NK and the $M_\mr{ej}$, those
ranges are rather arbitrary, due to the great uncertainty on the BH
formation process. Nevertheless, larger values for the NK and for the
$M_\mr{ej}$ than the extreme ones we chose, would lead to an unbinding
of the system, hence they are of no interest to our
work. 

Using the formalism we showed in Section~\ref{subsec:SN}, we compute $a_\mr{post}$ and $e_\mr{post}$. We assume that the SN ejecta
do not have any impact on the companion star properties, its spin for
instance, hence $\omega_{\star, \mr{post}}=\omega_{\star,
  \mr{pre}}=\omega_{\mr{orb,pre}}$. To account for the coupling
between tides and magnetic braking, we make sure that $a_\mr{post}$ is
less than the value of the maximal orbital separation at that
eccentricity and at the calculated $a_\mr{RLO}$ (Section~\ref{subsec:coupling}). An additional
criterion which allows to {{constrain}} the allowed parameter space of
NK and $M_\mr{ej}$ is the kinematical one: we make sure that
the peculiar velocity acquired by the binary at BH formation is larger
than the minimum peculiar velocity $v_{\perp,\mr{min}}$ we inferred from the position of
the binary (see Section~\ref{subsec:kin}).

\subsection{Observational biases on our sample}
\subsubsection{Results from the population model}

Because we want to derive general conclusions on the NK and ejected
mass at BH formation, we want to generalize the results for the
individual systems. In order to do so, we want to have some idea of
the effect of the different NKs and ejected masses on the observational
properties of the systems, and thus of possible observational biases
that make that the observed systems are predominantly those with
particular NKs and/or ejected masses. 

\begin{figure*}
\begin{minipage}{\columnwidth}
\centering
\includegraphics[width=\textwidth]{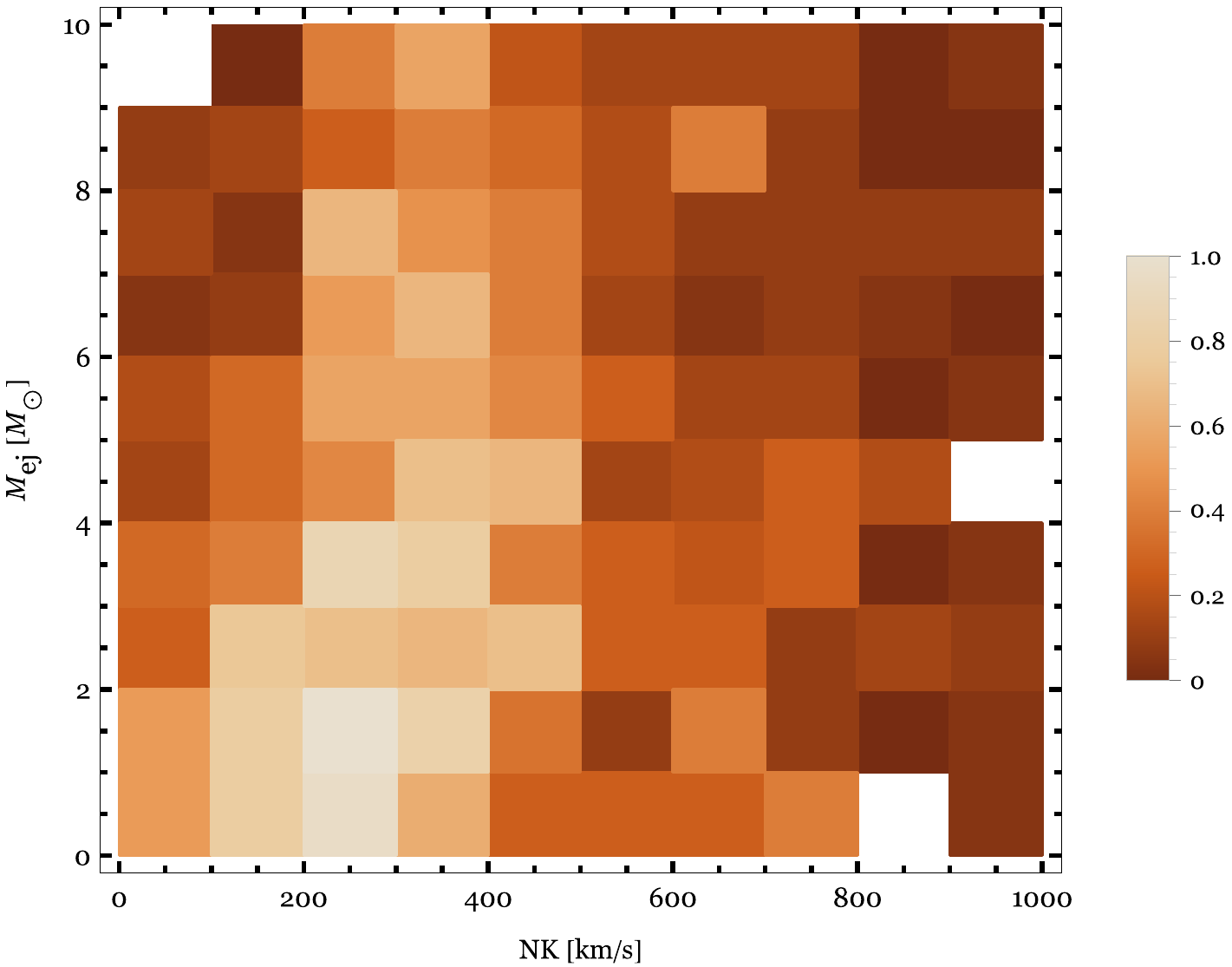}
\caption{Results of a binary population synthesis study of black-hole LMXBs. We select only those binaries which undergo RLO within the MS lifetime,
and we show their optimal parameter space for the mass ejected in the SN and the NK.}
\label{fig:BPS1}
\end{minipage}
\hspace{0.3cm}
\begin{minipage}{\columnwidth}
\centering
\includegraphics[width=0.95\textwidth]{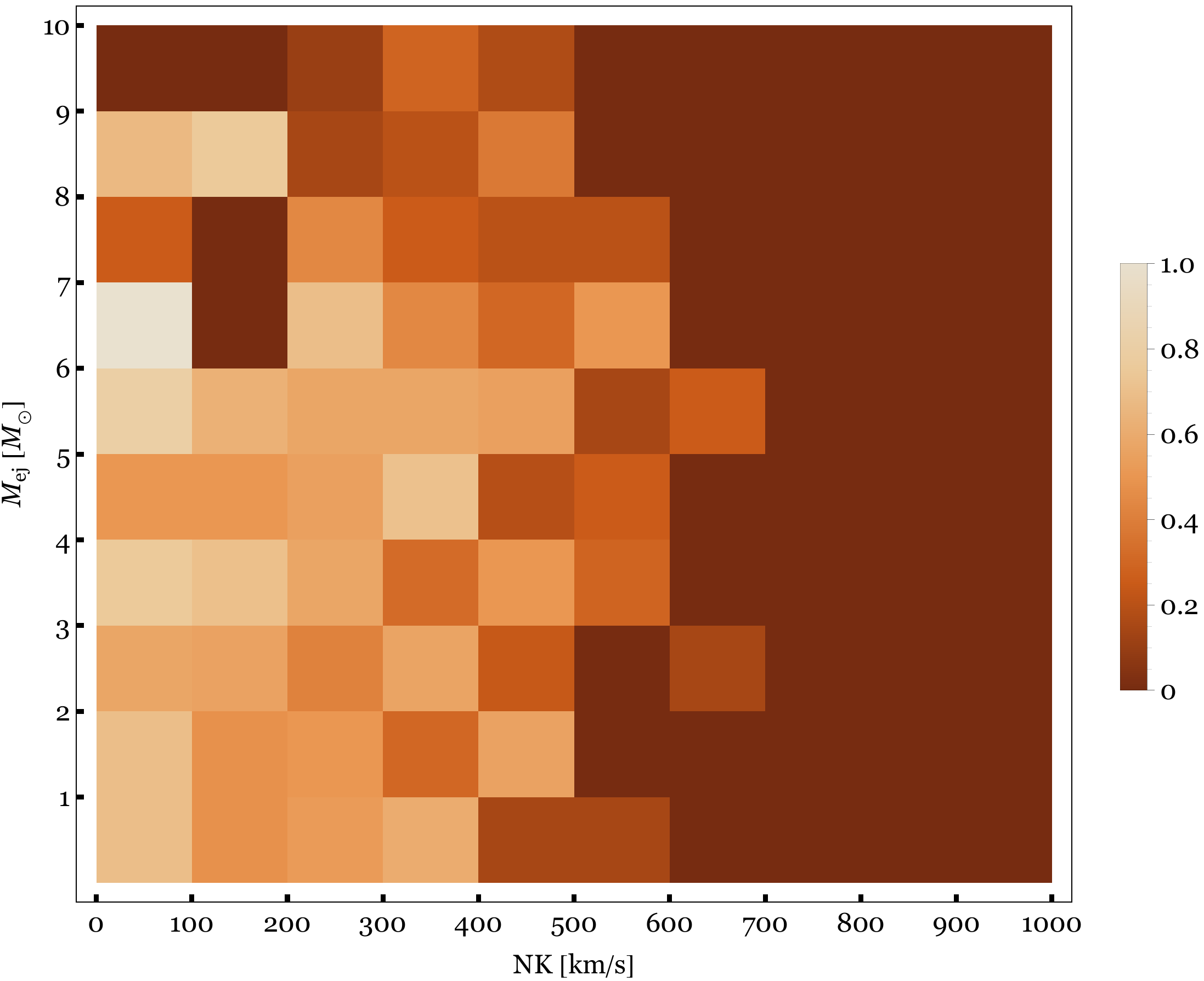}
\caption{For every 2D-bin of the left plot, we show here the fraction of systems which
reside within $10$ kpc from the Sun after orbiting in the Galactic potential for $10$ Gyr.}
\label{fig:BPS2}
\end{minipage}
\hspace{0.3cm}
\end{figure*}

We build a population of BH-LMXBs formed by a BH of mass $8~M_\odot$
and a companion of mass $1~M_\odot$. 
The magnitude of the NK, the mass ejected at BH formation, and the initial orbital separation 
are drawn from 
uniform distributions between $0$ - $1000$ km/s,
$0$ - $10~M_\odot$, and $0$ - $50~R_\odot$ respectively,
whereas the orientation of the NK is random over a sphere centered on the BH progenitor.
For those binaries that stay bound in the SN, we select only those ones which undergo RLO within the MS lifetime.
We show the results in Fig. \ref{fig:BPS1}. The higher density region
is for mild kicks around $100-200$ km/s,
and small $M_\mr{ej}$. Such a combination of these two parameters, allow the 
orbital separation in the post-SN configuration 
to fall within $a_\mr{max}$.

We then populate the Galaxy
following the stellar density in the thin disc (\citealt{2008gady.book.....B} and \citealt{2011MNRAS.414.2446M}), and we follow the
trajectories of the successful binaries for $10$
Gyr. The observed BH low-mass X-ray binaries all lie within $10$ kpc from the Sun (see Table \ref{tab:tabkin}).
This is very likely due to the fact that the binary has to be relatively close so that
a dynamical measurement of the BH mass via optical spectroscopy is possible.
We analyze how many of the simulated binaries satisfy this proximity-criterium 
as a function of the NK and $M_\mr{ej}$.
We show the results of this calculation in Fig. \ref{fig:BPS2},
where for every bin (NK, $M_\mr{ej}$) of the plot in Fig. \ref{fig:BPS1}, we compute the fraction of systems that lie within $10$ kpc from the Sun with respect to
the total number of winning binaries in that bin.
From the plot, it is evident that we 
are biased towards binaries
in which the BH received a low NK.
So even if BHs received high kicks,
we would not observe most of them.

\subsubsection{Hyper-velocity systems}
Related to the previous discussion,
we wonder how likely it is that we are missing systems in the halo of
the Galaxy.  Taking a Galactic local escape velocity of $\sim 500$
km/s (\citealt{2007MNRAS.379..755S}), the minimum NK the BH
has to acquire for the binary to escape the Galactic potential, is
$\sim 550$ km/s (for every choice of initial orbital separation and
ejected mass in the SN).  A binary moving with a peculiar
velocity of $500$ km/s would travel a distance of $500$ kpc in $1$
Gyr.  However, the probability for the binary to stay bound with such
a large NK is of $\approx 0.1$ only, as we can see in Fig.
\ref{fig:boundprob}.  Thus, we expect most of these hyper-velocity
binaries to get unbound.

Yet, it is likely that we are missing systems in the halo of our Galaxy.
We performed a test integrating the Galactic orbit of BH-LMXBs
in which the BH received a NK at birth. Even with a mild NK
of $200$ km/s, the BH-LMXB would move far away from the Sun reaching the halo in $10$ Gyr, out to $R\sim 60$ kpc.
So even if the binary stays bound in the BH formation event,
it would remain undetectable optically.

\section{Results}
\label{sec:results}
With the numerical method described above we calculate the possible combinations
of NK and ejected mass that are compatible with the observed
properties of the seven BH-LMXBs in our sample.

\subsection{Different models}
We vary the physics
involved in the binary evolution (mass-radius power-law index $\alpha$,
maximum transferred mass $\Delta M$) and we take into account the
uncertainty in the distance of the source, to test what is the
resulting uncertainty on the lower limits for the NK and the mass
ejected in the SN. Furthermore, we test the different MB prescriptions
(VZ and IT, see Section~\ref{subsec:coupling}). The models we test
are:\\
\begin{enumerate}
\item (standard model) VZ, $\Delta M = 1 M_\odot, \alpha = 0.82$ 
\item IT, $\Delta M = 1 M_\odot, \alpha = 0.82$ 
\item VZ, $\Delta M = 0.4 M_\odot, \alpha = 0.82$ 
\item IT, $\Delta M = 0.4 M_\odot, \alpha = 0.82$ 
\item VZ, $\Delta M = 1 M_\odot, \alpha = 0.69$ 
\end{enumerate}
For all the models, we
simulate $80\times 10^6$ initial binaries.

\subsection{Lower limits on the natal kick and ejected mass}

\begin{figure*}
\centering
\includegraphics[width=\textwidth]{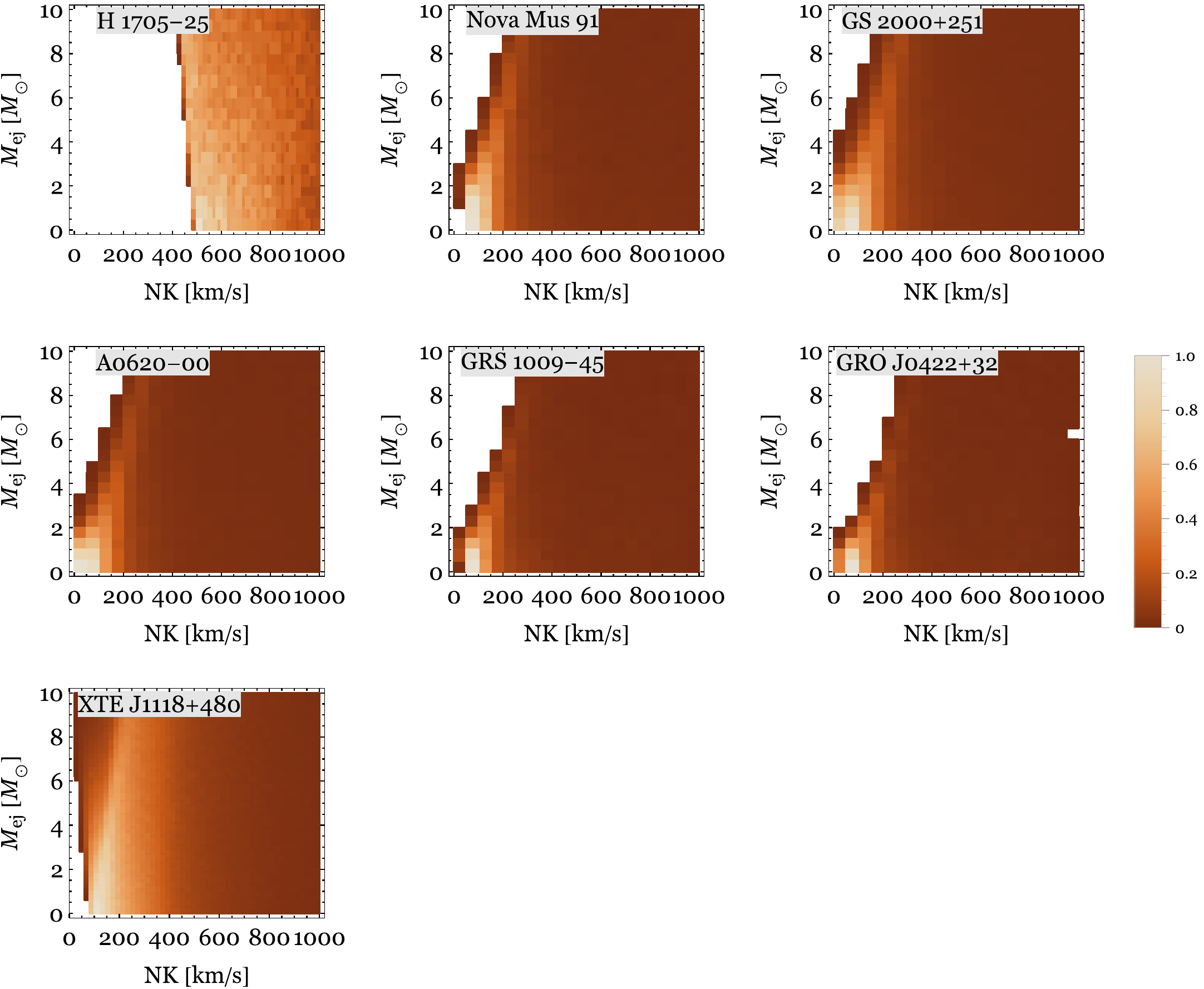}
\caption{Density plots showing the allowed parameter space for the
  mass ejected $M_\mr{ej}$ and the natal kick NK at BH formation for the seven
  short-period black-hole low-mass X-ray binaries,
  {{in the framework of our standard model (i),
  see text for details.}}
  }
\label{fig:density}
\end{figure*}

We show in Fig. \ref{fig:density} the density plots of possible
combinations of NK-$M_\mr{ej}$ that can lead to the currently observed
properties of the seven systems, for Model (i).

There is a typical trend in the higher-density region of the plots,
which indicates an increasing ejected mass for an increasing natal
kick.  This is caused by the constraint on the orbital separation in
the post BH-formation configuration: the binary has to be compact
enough for mass transfer to start while the companion is still on the
MS. In the absence of a large NK, large ejected mass would widen the
system too much. Only if a NK counters that effect are larger ejected
masses permitted;
but there is a decreasing fraction of NK-orientations 
that allow this
(see the decreasing density along the trend).

The density plots clearly separate the systems in at least two classes: five
systems that have the highest densities at or close to the origin,
and two systems (H1705-25 and XTE J1118+480) that have the highest
densities at large NK. However, there is still a significant
difference between the five low-kick systems: two of them (Nova Mus
91 and GRS 1009-45) actually exclude or have very low density to
have no kick. Finally, we note that all systems have significant
density at $M_\mr{ej} = 0$. 

We need to point out that these distributions do not
correspond to probability distributions of the actual values of the
parameters for the systems, as the parameter range and parameter
distributions that we used in the Monte Carlo (flat) will likely not
represent the real ones. However, as we used flat input distributions,
the resulting probabilities do show how much fine tuning in the
initial binary parameters and NK orientations is needed to survive and match
the kinematic and orbital properties of the binaries.

\begin{figure*}
\begin{minipage}{\columnwidth}
\centering
\includegraphics[width=0.9\textwidth]{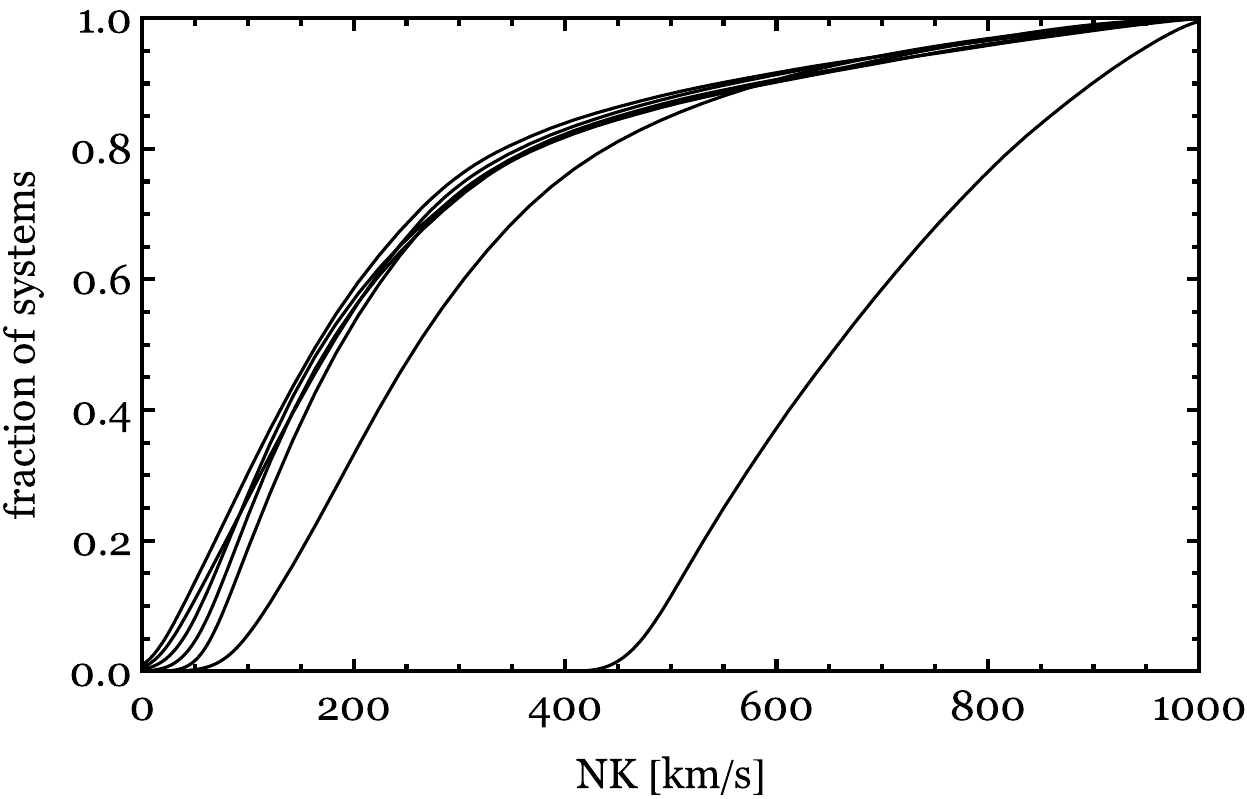}
\caption{Cumulative distribution for the natal kick received by the black hole
  in short-period BH-LMXBs,  {{in the framework of our standard model (i),
  see Text for details}}. From left to right: 1A 0620-00, GS
  2000+251, GRO J0422+32, GRS 1009-45, Nova Mus 1991, XTE J1118+480, H 1705-250.
  }
\label{fig:NKcum}
\end{minipage}
\hspace{0.3cm}
\begin{minipage}{\columnwidth}
\centering
\includegraphics[width=0.9\textwidth]{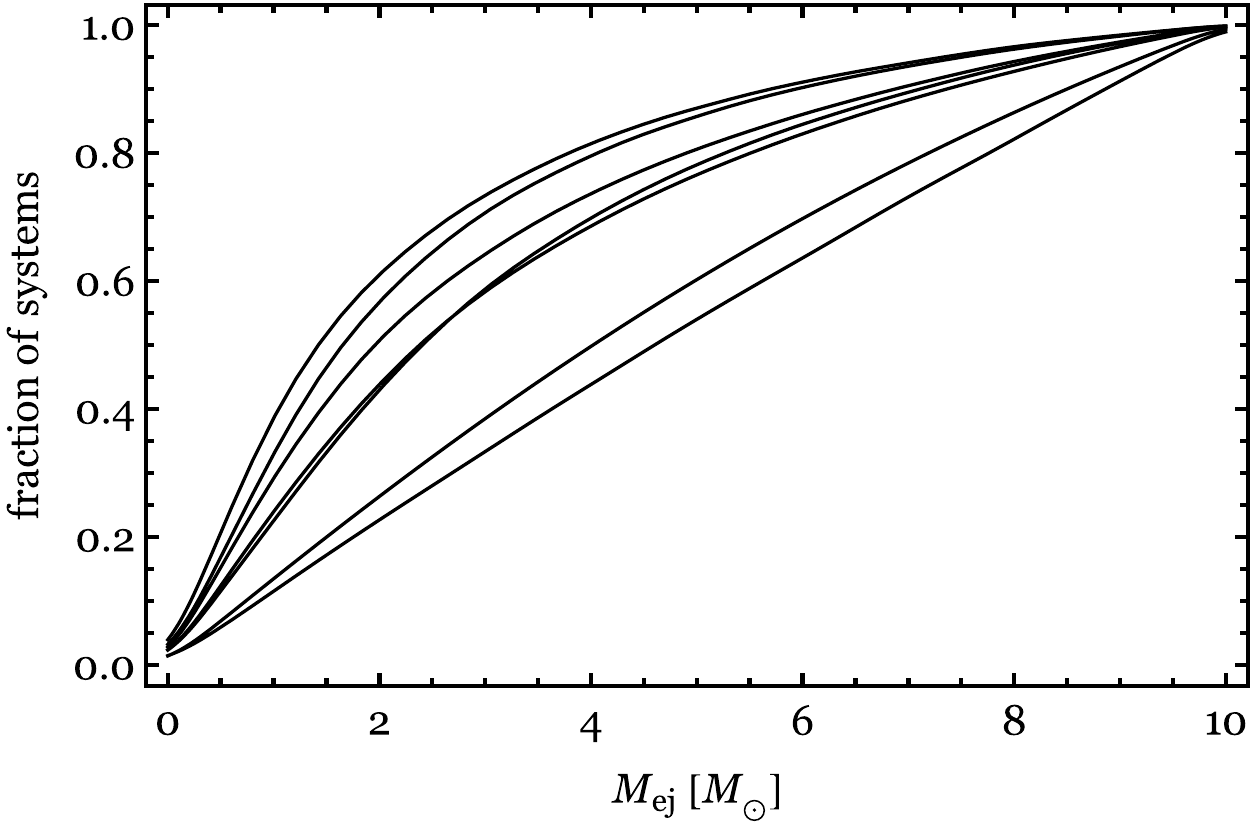}
\caption{Cumulative distribution for the ejected mass at BH formation
  in short-period BH-LMXBs,  {{in the framework of our standard model (i),
  see Text for details}}. From left to right: GRO J0422+32, 1A
  0620-00, GRS 1009-45 and Nova Mus 1991 (overlapping curves), GS
  2000+251, XTE J1118+480, H 1705-250.}
\label{fig:Mejcum}
\end{minipage}
\hspace{0.3cm}
\end{figure*}

In order to quantify our results, we show in Fig. \ref{fig:NKcum}
and \ref{fig:Mejcum} the cumulative distribution for the NK and for
the ejected mass for each of the $7$ sources for our Monte Carlo
calculations. 
We define a \emph{minimum} NK and ejected mass as a
  cut at the $5\%$ probability in these cumulative distributions, i.e.  a
  limit at which the $95\%$ of the cumulative distribution is
  higher. These lower limits are shown in Table \ref{tab:min} and the range
  of values for each of the sources is a consequence of the
  uncertainty on the physics involved (i.e. the different models we use) and on the distance. 
  The variation of the parameters involved in the binary evolution,
  namely the amount of mass transferred to the BH,
  the mass-radius index and the strength of MB,
  do not affect greatly the lower limits on the NK and on the $M_\mr{ej}$
  (see second and third column in Table \ref{tab:min} for the first six binaries). 
  These lower limits are mostly affected by the peculiar velocity of the system,
  hence by the height of the system from the Galactic plane,
  and are therefore consistent with the ones by \citet{2012MNRAS.425.2799R}
that accounted for the kinematics of the sources only.
 Indeed, in the case
  of H 1705-250, the large range for the NK is due to the large
  ($25\%$) uncertainty in the distance to the source. Whereas the large value for the NK
is due to 
  the fact that the binary resides right above the Galactic bulge, very close to the Galactic centre where the Galactic potential is
  strongest. 
  
  Typically, these lower limits fall in
  lower density regions of the plots in Fig. \ref{fig:density} and
  thus still require severe fine tuning of in particular the NK orientation.
  We also wish to note that assuming all (x, y, z) directions for the NK are equally probable,
  the lower limits for the NK indicated in the table are to be multiplied
  by the square root of $3$.
  
  We also checked the effect of taking a non-conservative mass transfer rather than a conservative one, using
  formula \ref{eq:MBNCMT} for tracing backwards the semi-detached phase,
  and taking $\beta=0.1$. When comparing the results of this model with the results of Model (i),
  for example, we find that the lower limits on the NK and on $M_\mr{ej}$
  do not change by more than $7\%$,
  for all the $7$ sources.

\begin{figure}
\centering
\includegraphics[width=0.9\columnwidth]{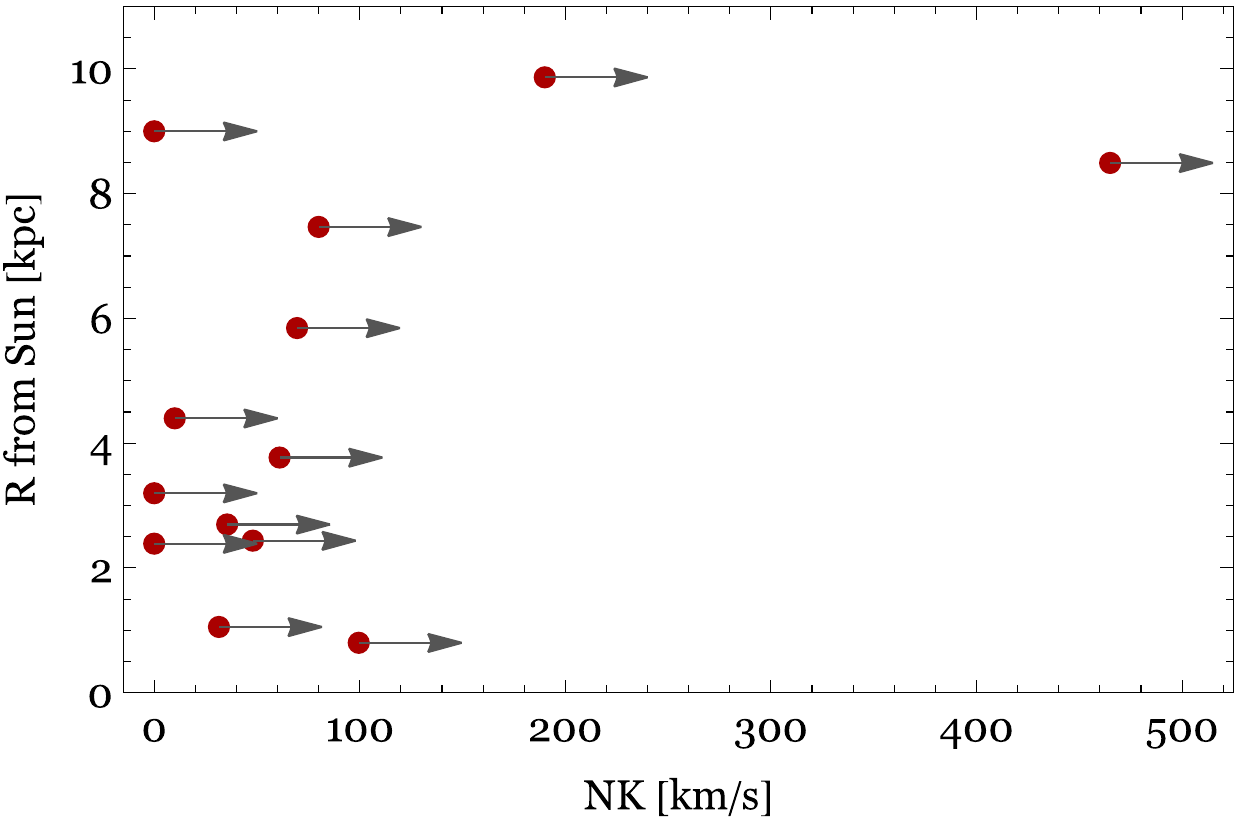}
\caption{BH low-mass X-ray binaries as a function of their distance $R$ from the Sun and the lower limit on the NK,
as computed in Paper I and in this work.}
\label{fig:RsunNKobs}
\end{figure}

The fact that all the highest-density regions in Fig. \ref{fig:density} are skewed towards low kicks
(except for H 1705-250 and XTE J1118+480),
does not mean that high BH kicks are excluded. 
The reason why we are biased towards small NKs
is the proximity of the sources to the Sun.
This can be seen in Fig. \ref{fig:RsunNKobs},
where we plot the whole sample of confirmed BH low-mass X-ray binaries  
as a function of their distance to the Sun and the lower limit on the NK.
Referring back to our BPS study, it is evident
that the higher the kick, the lower the probability for the system to be in the proximity of the Sun 
(see Fig. \ref{fig:BPS2}).\\

From our results, it is possible to highlight three different cases:
\begin{enumerate}
\item {\emph{Standard systems}}: systems with a low peculiar velocity at birth, which can be
  consistent either with a small NK, or with a kick imparted to the binary as a result of the mass ejection in the SN event. 
\item {\emph{Zero ejected mass systems}}: systems consistent with a non-zero NK {\emph{but}} zero ejected
  mass at BH formation.
\item {\emph{High NK systems}}: systems consistent with a high NK,
comparable to NS natal kicks.
\end{enumerate}
We will discuss these three scenarios in Sec. \ref{sec:discussion}.

\begin{table*}
\caption{The second and third column show the minimum natal kick and the minimum mass ejected at BH formation. The third column shows $a_\mr{pre,max}$ such that the binary undergoes mass transfer within the MS lifetime. 
The range indicates how the values vary when changing the assumption 
on the parameters involved in the binary evolution of the sources (see Text for a description of the different models we consider). The last column shows $a_\mr{pre,max}$ such that the binary stays bound in the SN.}
\label{tab:min}
\begin{tabular}{| c | c   | c | c | c |}
\hline
{\hspace{0.5cm}{Source}}\hfill &{{min NK}} & min $M_\mr{ej}$& {{max $a_\mr{pre}$, RLO on MS}} &{{max $a_\mr{pre}$}}, bound in SN \hfill\\
& \small{[km/s]} & \small{[$M_\odot$]} & \small{[$R_\odot$]} & \small{[$R_\odot$]} \\
\hline\hline
GS 2000+251 & 24-47 & 0.13-0.33 & 9-37 & 7800\\
A0620-00 & 20-43 & 0.09-0.32 & 8-37 & 8400\\
Nova Mus 91 & 62-77 & 0.17-0.34 & 8 &  1400\\
XTE J1118+480 &  93-106 & 0.31-0.37  & 23-38 & 570\\
GRS 1009-45 &  49-73 & 0.08-0.28 &  8-38 & 2400\\
GRO J0422+32 & 35-61 & 0.04-0.26 &7-38 &  3000\\
H 1705-250  & 415-515 & 0.40-0.50 & 11-19 & 27\\
\hline
\end{tabular}
\end{table*}

\subsection{Correlation natal kick VS mass of the BH}
\begin{figure}
\centering
\includegraphics[width=0.9\columnwidth]{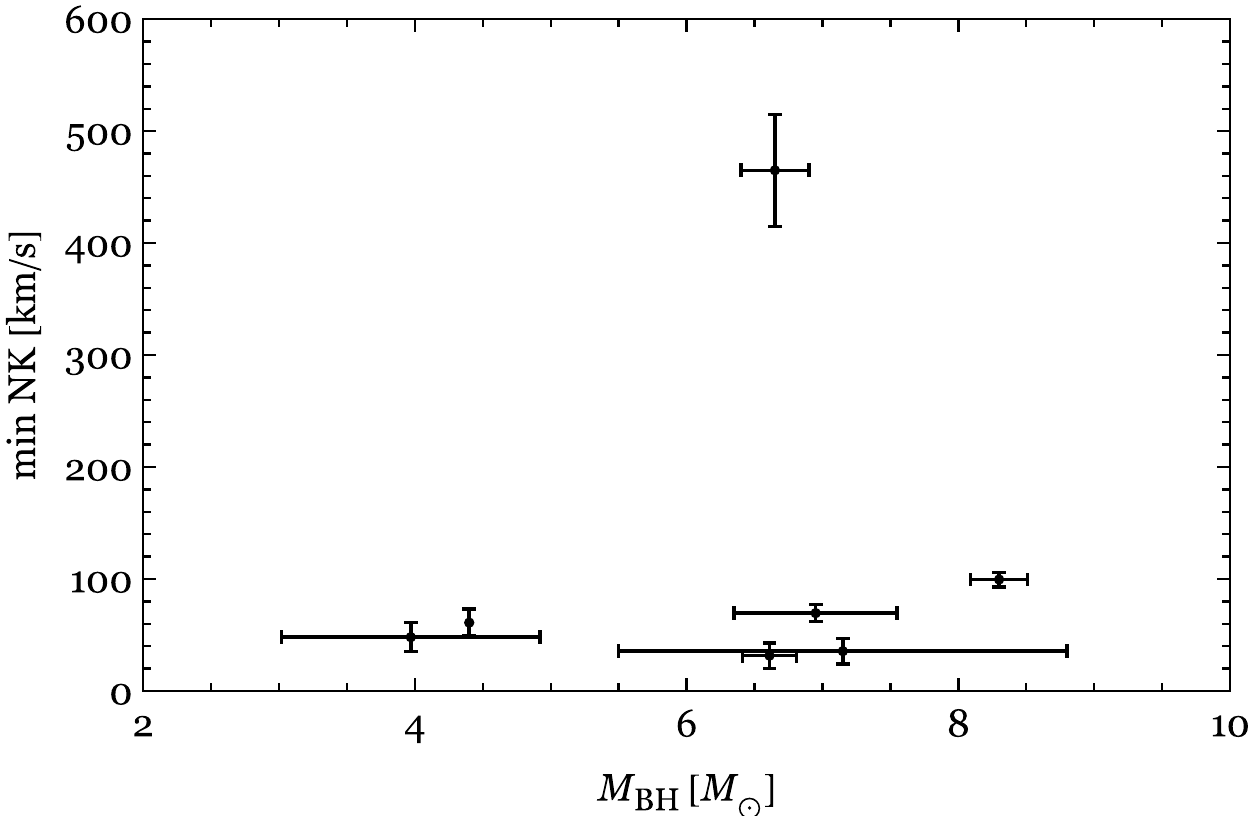}
\caption{Short-period BH low-mass X-ray binaries as a function of black hole mass
and the minimum natal kick which results from our simulations.}
\label{fig:correlation}
\end{figure}

In Fig. \ref{fig:correlation} we plot our lower limits for the NK as
a function of the BH mass to see if there is any
correlation. From our results, we find so far (with our limited sample) no evidence for a correlation
between NK and BH mass. If we found a correlation,
such as lighter BHs receiving higher kicks,
that could be a hint for
a NK happening on a shorter timescale than the fallback timescale.

\subsection{Initial orbital separation}

\begin{figure*}
\begin{center}
\includegraphics[width=1.0\textwidth]{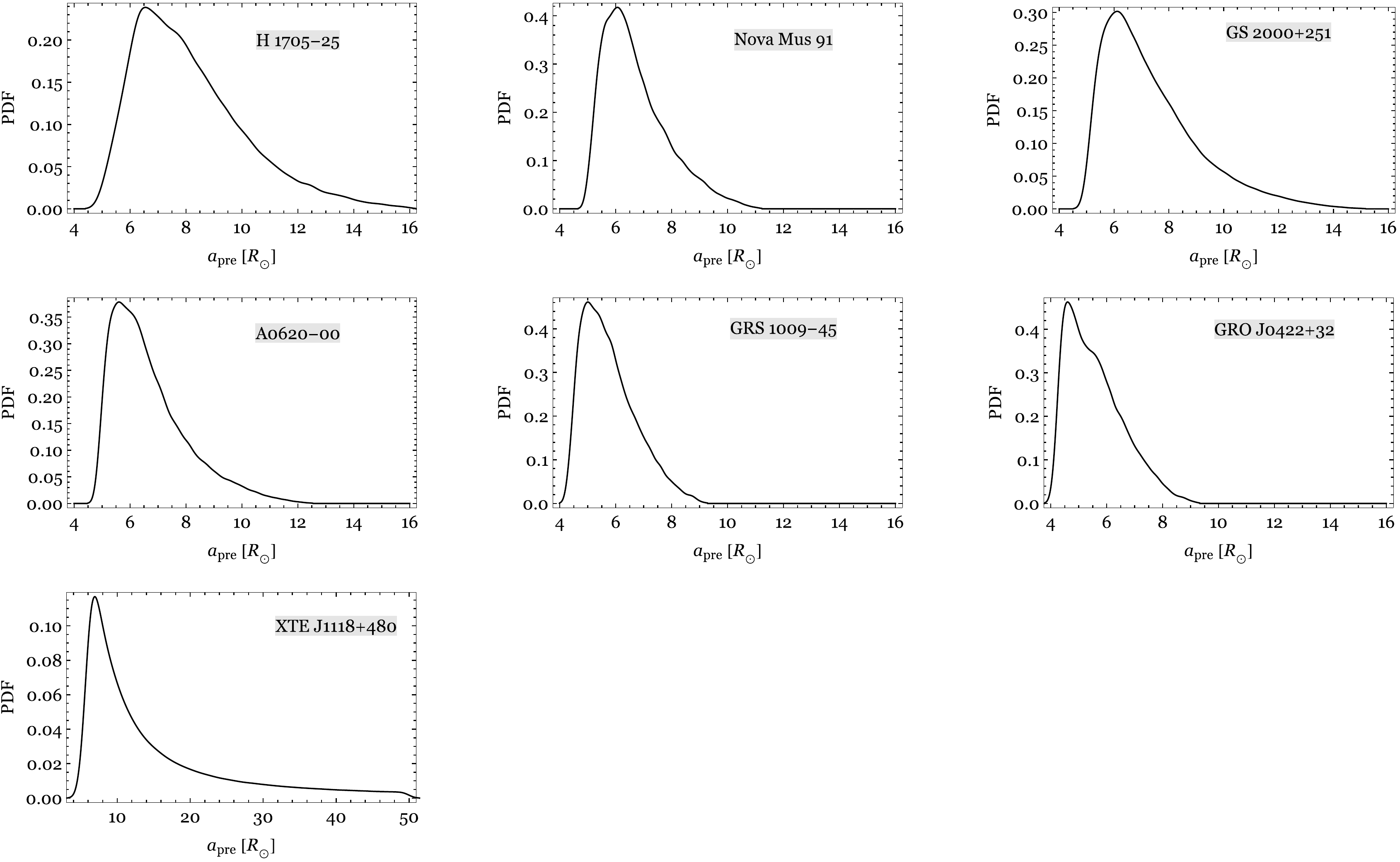}
\caption{Probability density function for the orbital separation right
  before black hole formation for short-period BH-LMXBs,
    {{in the framework of our standard model (i),
  see Text for details}}.
  }
\label{fig:aprepre}
\end{center}
\end{figure*}

We show in Fig. \ref{fig:aprepre} the probability density function
for the orbital separation right before BH formation
for the seven sources.

In the third column of Table \ref{tab:min}, we show the maximal
orbital separation right before BH formation such that the binary
stays bound in the SN {\emph{and}} undergoes mass transfer within the
MS-lifetime. In the fourth column, we show the maximum value
for $a_\mr{pre}$ such that the binary stays bound in the SN. The range
of values correspond to varying our models. Comparing the values of the last two columns,
we see that the constraint to have RLO within MS lifetime,
is much more limiting than having the binary to stay bound in the SN.
As a consequence, we expect that there are many detached BH+MS-star binaries
which would evolve to longer and longer periods due to the nuclear expansion of the companion,
finally evolving into BH+WD binaries.

\section{Discussion}
\label{sec:discussion}

There are three BH-formation scenarios  which are compatible with some of the seven sources we studied.

\subsection{Zero ejected mass systems}
Five systems (H 1705-250, Nova Mus 91, 
GRS 1009-45, GRO J0422+32, XTE J1118+480) 
are consistent with a NK, while
not requiring the ejection of mass at BH formation. 
The NK at which the one dimensional NK density function for $M_\mr{ej}=0$ peaks
is 
correspondingly: $525$, $76$, $68$, $54$, $114$ km/s.
We wish to stress that these value do not correspond to the most likely value of a probability density function, rather
they are the values of the NK which require less fine-tuning.

Zero ejected mass at BH formation could be
consistent with a BH formed in the dark (no SN, i.e. no baryonic mass ejected) and with a
NK caused by asymmetric neutrino emission or asymmetric GW
emission.  When a BH is formed,
the gravitational mass defect which is equal to the negative binding energy,
is calculated to be $\sim 10\%$ of BH mass (Zeldovich \& Novikov 1971).
If this mass leaves the system in the form of neutrinos,
the predicted ejected mass can be consistent with our limits of Table \ref{tab:min}.

We note, however, that in the case of XTE J1118+480
there is evidence for an explosion having occurred from
the chemical enrichment in the spectra of the companion to the BH
(\citealt{2006ApJ...644L..49G}).

Furthermore, we wish to highlight that our study
is the first in suggesting 
that few of the BHs in LMXBs might have been formed without baryonic mass ejection.
This was found for few BHs in high-mass X-ray binaries
(see \citealt{2010Natur.468...77V}, \citealt{2003Sci...300.1119M}).

\subsection{Standard systems}
\begin{figure*}
\begin{center}
\includegraphics[width=\textwidth]{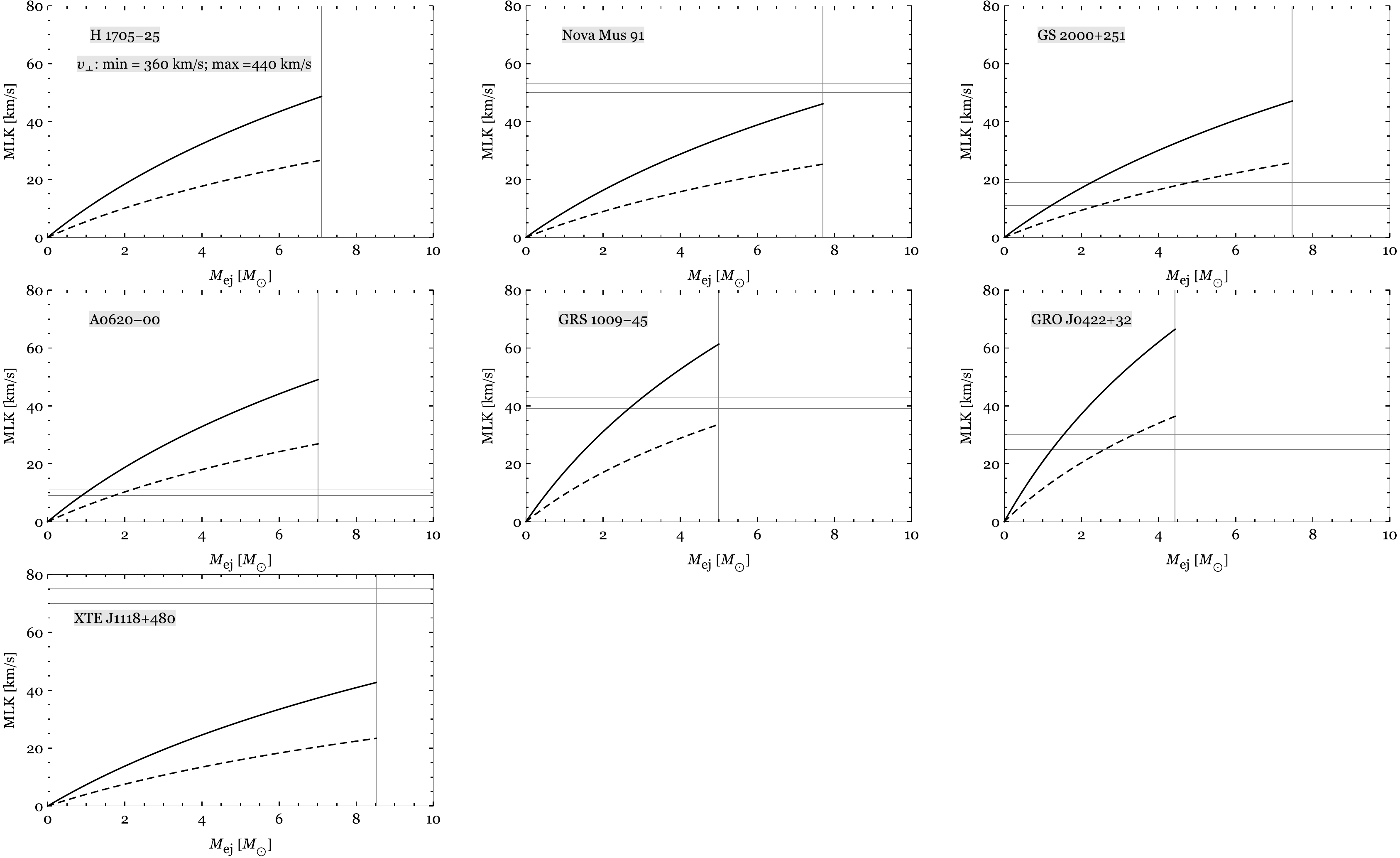}
\caption{Mass-loss kick (MLK) as a function of the mass ejected at BH formation
for the short-period BH-LMXBs. {{The vertical lines show the upper limit on the mass ejected at BH formation for the binary to stay bound.}} The solid gray lines correspond to the lower and upper limit on the peculiar velocity at birth. The black solid and black dashed line correspond to different assumptions on the pre-SN orbit (see Text). {{The point of intersection between these lines
and the limits on the peculiar velocity
depicts when a system can be explained by mass ejection only.}}}
\label{fig:niceplots}
\end{center}
\end{figure*}

Five sources out of seven (Nova Mus 91, GS 2000+251, 1A 0620-00, GRS 1009-45, GRO J0422+32)
require only a small
peculiar velocity at birth, of the order of few tens of km/s or less
(see Table \ref{tab:tabkin}).  This small peculiar velocity could be
imparted to the binary as a result of mass ejection in the SN event.
We can compute how much ejected mass is needed to account for the
small peculiar velocity (see Fig. \ref{fig:niceplots}),
in a similar manner as done previously by
\citet{1999A&A...352L..87N}.
In order to compute the kick received by the binary due to the mass ejection (MLK),
we use formula \ref{eq:massloss}. We trace the orbital binary properties backwards assuming
a transferred mass of $1~M_\odot$ and $\alpha=0.82$. As a test,
we assume a semi-major axis in the pre-SN configuration of $1.5~a_\mr{RLO}$ (solid line) and
of $5~a_\mr{RLO}$ (dashed line). The plots show the amount of mass ejected needed to account for the peculiar velocity of the $5$ sources aforementioned. 
There is an upper limit on $M_\mr{ej}$ such that the binary stays bound in the SN,
i.e. the ejected mass has to be less than half of the total initial mass,
which translates into $M_\mr{ej}<M_\mr{BH}+M_\star$ {{(vertical lines in Fig. \ref{fig:niceplots}).}}
All of the five systems but Nova Mus 91,
can be explained by mass-ejection only.

Alternatively,
density wave scatterings could impart a velocity to the system of the order of few tens km/s
(\citealt{1995MNRAS.277L..35B}),
with a maximum value of $40$ km/s (\citealt{1981gask.book.....M}).

{{As previously discussed in Sec. \ref{sec:OrbProp}, 
GRS 1009-45 lacks strong constraints on the masses of its components.
In particular, the BH mass quoted in Table \ref{tab:tabOBSpro}
is a strict lower limit.
We then estimate the MLK taking a larger BH mass
of $7~M_\odot$
and a companion mass of $0.98~M_\odot$.
We find that for $a_\mr{pre}=1.5~a_\mr{RLO}$,
the ejected mass needed to account for the peculiar velocity of the system
is $\sim 7~M_\odot$. Calculations by \citet{2001ApJ...554..548F}
indicate a maximum ejected mass at BH formation of $\sim 4~M_\odot$.
We conclude that it is essential to better constraint the component masses in 
GRS 1009-45 in order to discriminate between a standard scenario and
a non-zero NK scenario. 

Concerning GRO J0422+32,
we check what is the effect of 
using BH and companion-star masses from 
\citet{2007MNRAS.374..657R}
($M_\mr{BH}=10.4~M_\odot$ and $M_\star=0.4~M_\odot$).
For $a_\mr{pre}=1.5~a_\mr{RLO}$,
the MLK is always lower than the peculiar velocity of the systems,
for every $M_\mr{ej}$. Again, this highlights the importance of correctly 
estimating the component masses.}}

\subsection{High natal kick systems}
Two out of seven sources require a high NK at
birth (H 1705-250 and XTE J1118+480),
comparable to NS natal kicks. This could be a hint for the NK happening
on the same timescale of the fallback.
A high NK for the BH in XTE J1118+480
was already suggested by
\citet{2009ApJ...697.1057F}.

As an alternative explanation, these two systems could have been ejected from a GC.
None of the BH X-ray binaries
currently known are found in a GC. So far, there
is evidence for two BHs in a Milky Way GC
(\citealt{2012Natur.490...71S}), and one BH in an extragalactic GC
(\citealt{2007Natur.445..183M}).  However, no optical companion
has yet been found for these candidates,
hence an accurate estimate of the mass of the compact object does not exist.
A GC origin for XTE J1118+480 seems unlikely,
due to the super-solar surface metallicity of the companion
(\citealt{2009ApJ...697.1057F}, \citealt{2006ApJ...644L..49G}).

H 1705-250 is the source which looks most different
from all the other sources in the sample. It is located right above the Galactic bulge,
at Galactic coordinates $R\sim 0.5$ kpc, $z\sim 1.35$ kpc.
Its systemic radial velocity was measured by \citet{1996ApJ...459..226R},
giving a value of $10\pm 20$ km/s.
We conclude that either the system is moving perpendicularly to our line of sight,
or that a GC origin for the system is more likely.

Looking at Fig. \ref{fig:plottrattini},
there might be at least $5$ other systems
hinting for a high NK, those ones whose peculiar velocity at birth $v_{\perp,\mr{min}}$ is
similar to the one of XTE J1118+480. Future work is needed to precisely determine the mass of the compact object in such systems.

\subsection{Effect of the uncertainty on the distance}
{{The most accurate method to determine the distance to a BH
X-ray binary, is to estimate the absolute magnitude of the companion star, 
see \citet{2004MNRAS.354..355J}.
This method has been used for all the 7 BH-LMXBs in our sample,
except for Nova Mus 91,
for which the distance was estimated by {{\citet{2005ApJ...623.1026H}}}
through the interstellar absorption properties of the source,
which is a less reliable method.
{{\citet{1996ApJ...468..380O}}} found a distance of $d=5.5\pm 1$ kpc,
estimating the spectral type and luminosity of the companion.
Such a distance would give $v_{\perp\mr{,min}}=50-56$ km/s,
still consistent with our value of $v_{\perp\mr{,min}}=50-53$ km/s (see Table \ref{tab:tabkin}).

The major source of systematic errors when 
estimating the distance through the absolute magnitude of the companion,
comes from a failure in accounting for the right disc-contribution to the light of the star
(see Sec. 6 in Paper I for a discussion).
{{\citet{2005ApJ...623.1026H}}} warn about possible systematics on the measurement
by {{\citet{2001PhDT..........G}}} of the distance to GRS 1009-45, due to the assumed small veiling from the disc.
Taking into account a larger contribution 
from the disc, {{\citet{2000ApJ...533..329B}}}
obtained a distance $d=5-6.5$ kpc.
This translates into a minimum peculiar velocity
$v_{\perp\mr{,min}}=48-52$ km/s,
larger than the one taken in our work (see Table \ref{tab:tabkin}).
This results in a minimum NK of $\sim 62$ km/s,
when performing a set of simulation in the framework of our standard model (i).
This value is still consistent with the range shown in Table \ref{tab:min}.
}}

\section{Conclusions}
There are five main conclusions which result from this work:
\begin{enumerate}
\item The lower limit on the NK is not greatly affected by the binary evolution of the
sources.
It is mostly affected by the kinematics of the system,
and therefore by the uncertainty on the distance. In this respect,
our results are consistent with the ones in \citet{2012MNRAS.425.2799R},
who calculated lower limits on BH natal kicks basing their study on kinematical arguments.
Variations of the assumptions on MB and mass transfer give very similar results.
\item Even if the lower limit on the NK is not affected by the binary evolution of the system,
in order to unravel what are the optimal combinations of NK
and mass ejected in the SN, it is necessary to follow the details of the whole evolutionary paths of BH-LMXBs. In particular, this method allowed us to find binaries consistent with a neutrino-driven NK.
\item Our work enables us to highlight three possible scenarios for the birth of the BH.
Two of these scenarios have been discussed previously in the literature:
either the BH does not receive any NK,
or it receives a NS-like NK. The third scenario that we suggest,
is a BH having formed with a NK and zero baryonic mass ejection. It is the first time that this scenario has been applied to the evolution of
BH-LMXBs, whereas it was first found to be consistent with the formation of few BHs in high mass X-ray binaries
(see \citealt{2010Natur.468...77V}, \citealt{2003Sci...300.1119M}).
\item We find evidence for a high NK (i.e. a NS-like NK) in two of the sources,
and potentially $5$ BH X-ray binary candidates whose 
minimum peculiar velocity at birth suggests a high NK.
\item {{Our population study highlights that,
due to the limits of optical spectroscopy,
there exists a bias towards BH-LMXBs being close (within $10$ kpc) to the Sun.}}
For the same reason,
NK estimates are biased towards low/mild NKs (less than $100-200$ km/s).
\end{enumerate}

\section{Acknowledgments}
We are grateful to the anonymous referee for very pertinent insights,
especially on the observational properties of the binaries,
and on the implications of the uncertainty on these properties on our study.
SR thanks Bailey Tetarenko for having provided the catalogue 
of BH candidates, and for her thorough
and critical reading of the manuscript.
The work of SR is supported by the Netherlands Research School for
Astronomy (NOVA).

\appendix

\section{Analytical treatment of the mass transfer phase}
\label{sec:appA}
The orbital angular momentum of a binary is written as
$$J_\mr{orb}=\mu\sqrt{G M a},$$
in terms of the reduced mass $\mu$, the total mass of the binary $M$, and the semi-major axis $a$.
Using \citet{1971ARA&A...9..183P} relation for the Roche radius, $R_L\approx 0.46~a~\left ({\frac{M_\star}{M}}\right )^{1/3}$, 
and the fact that the star is filling its Roche lobe during mass-transfer, $R_L=R_\star=fM_\star^\alpha$,
we can express $J_\mr{orb}$ in terms of the component masses only:
$$J_\mr{orb}=\sqrt{Gf}M_\mr{BH} M^{-1/3} M_\star^{\frac{5}{6}+\frac{\alpha}{2}}$$\\
This expression can then be derived with respect to $M_\star$:
$$\frac{dJ_\mr{orb}}{dM_\star}=M_\mr{BH} M^{-\frac{1}{3}} (\frac{5}{6}+\frac{\alpha}{2})M_\star^{-\frac{1}{6}+\frac{\alpha}{2}}\sqrt{Gf}-M^{-\frac{1}{3}}M_\star^{\frac{5}{6}+\frac{\alpha}{2}}\sqrt{Gf}$$
The balance equation \ref{eq:balance} then becomes:\\
$$\frac{\dot{a}}{a}=\frac{2}{J_\mr{orb}}\frac{dJ_\mr{orb}}{dM_\star}\dot{M_\star}+2\beta\frac{\dot{M_\star}}{M_\mr{BH}}-2\frac{\dot{M_\star}}{M_\star}+(1-\beta)\frac{\dot{M_\star}}{M}$$
The last equation can be integrated analytically obtaining
\ref{eq:MBCMT} and \ref{eq:MBNCMT}.

\section{Fitting formulae for the maximal orbital separation}
We show in Table \ref{tab:tabFitting} the parameters for the fitting of the maximal orbital separation as a function of the eccentricity (see expression
\ref{eq:fitamax}).
\label{sec:appC}
\begin{table}
\caption{{Fitting parameters for $a_\mr{max}$.}}
\label{tab:tabFitting}
\begin{tabular}{l c c c c c}
\hline
Source & $\alpha$ & $\beta$& $\gamma$ &$\delta$ & $\epsilon$ \\
\hline\hline
H 1705-250 & & & & & \\ 
model1 & 8.94 & -2.84 & -2.43 & -3.16 & 1.00\\
model 2 & 8.93 & -2.81 & -2.45 &-3.16 &1.00\\
model 3 & 11.00 & -3.38 & -2.40 & -6.28 & 1.00\\
model 4 & 8.60 & -4.70 & -0.90 & -4.63 & 1.00\\
model 5 & 7.84 & -1.68 & -3.94 & -3.16 & 1.00 \\
\hline
Nova Mus 91 &&&&&\\  
model 1 & 6.69 & -0.79 &-7.90 &-3.23 & 1.00\\
model 2 & 6.68 & -0.82 & -7.24 & -3.20 & 1.00\\
model 3 & 11.34 & -7.45 & -0.79 & -6.32 & 1.00\\
model 4 & 8.20 & -5.51 & -0.61 & -4.25 & 1.00\\
model 5 & 6.46 & -0.68 & -9.17 & -3.23 & 1.00\\
\hline
GS 2000+251 &&&&&\\ 
model1  & 8.22 & -1.96 & -3.49 & -3.20& 1.00\\
model 2  & 8.21 &  -1.94 & -3.47 & -3.18 & 1.00\\
model 3  & 11.01 & -2.70 &-3.38 & -6.23 & 1.00\\
model 4  &7.86 & -3.69 &-1.11 & -4.25 & 1.00\\
model 5 & 7.02 & -1.01 & -6.26 & -3.20 & 1.00\\
\hline
1A 0620-00&&&&&\\  
model 1 &  7.01 & -1.07 &  -5.86 &  -3.14 & 1.00 \\
model 2  & 7.00 & -1.07 & -5.68 & -3.12 & 1.00\\
model 3 & 10.80 & -3.05 & -2.73 & -6.16& 1.00\\
model 4  & 7.35 & -4.22 & -0.84 & -3.96 & 1.00\\
model 5  & 6.37 & -0.69 & -8.77 & -3.15 & 1.00\\
\hline
GRS 1009-45&&&&&\\  
model 1&  5.52 & -0.53 & -10.64 & -2.86 & 1.00\\
model 2 & 5.50 & -0.54 & -9.99 & -2.83 & 1.00\\
model 3 & 10.15 & -3.95  & -1.68 & -5.79 & 1.00\\
model 4  & 6.69 & -5.39 & -0.54 & -3.57 & 1.00\\
model 5 &  11.66 & -11.20 & -0.47  & -6.36 & 1.00\\
\hline

GRO J0422+32&&&&&\\ 
model 1 &  11.73 & -11.29 & -0.47 & -6.39 & 1.00\\
model 2 &  5.49 & -0.54 & -10.00 & -2.83 & 1.00\\
model 3 &  9.91 & -2.92 & -2.55 & -5.66 & 1.00\\
model 4 & 6.31 & -3.93 & -0.76 & -3.41 & 1.00\\
model 5 & 5.20 & -0.50 & -10.74 & -2.75 & 1.00\\
\hline
XTE J1118+480&&&&&\\  
model 1 & 11.73 & -11.29 & -0.47 & -6.39 & 1.00\\
model 2 & 8.68 & -4.09 & -0.90 & -4.43 & 1.00\\
model 3 & 10.98 & -2.25 & -4.41  & -6.11 & 1.00\\
model 4 & 6.67 & -3.09 & -1.14 & -3.62 & 1.00\\
model 5 & 11.66  & -11.20 & -0.47  & -6.36 & 1.00\\

\hline \\
\end{tabular}
\newline
\end{table}

\end{document}